\documentclass[english,aps,prb,superscriptaddress,reprint,longbibliograhy]{revtex4-2}

\usepackage[T1]{fontenc}
\usepackage{amsmath}
\usepackage{amssymb}
\usepackage{amsfonts}
\usepackage{bbold}
\usepackage{tikz}
\usepackage{float}
\usepackage{xfp}
\usepackage{physics}

\usepackage{color}

\usepackage{graphicx}
\usepackage{dcolumn}
\usepackage{bm}
\usepackage{commath}
\usepackage{upgreek}
\usepackage{physics}
\usepackage{natbib}

\usepackage[english]{babel}
\usepackage[colorlinks=true,linkcolor=cyan,citecolor=blue,urlcolor=cyan]{hyperref}
\usepackage[bottom=2cm,top=2cm, left=1.5cm, right=1.5cm]{geometry}
\usepackage{booktabs}
\usepackage{comment,verbatim}

\usepackage[caption=false]{subfig}

\colorlet{linkequation}{blue}
\usepackage[normalem]{ulem}

\newcommand{\showcomments}{\long\def\comm##1\commend{##1}}

\showcomments

\newcommand{\akcom}[1]{{\comm \color{red} Akash: ``#1'' \commend}}
\newcommand{\tmcom}[1]{{\comm \color{blue} Tarek-Moussa: ``#1'' \commend}}

\newcommand{\ak}[1]{{\color{black}#1}}

\begin{document}

	\title{Unquenched orbital angular momentum as the origin of spin inertia}
	\author{Tarek Moussa}
    \email{tmoussa@rptu.de}
	\affiliation{Department of Physics and Research Center OPTIMAS, Rheinland-Pf\"alzische Technische Universit\"at Kaiserslautern-Landau, 67663 Kaiserslautern, Germany}

	\author{Darpa Narayan Basu}
	\affiliation{Department of Physics and Research Center OPTIMAS, Rheinland-Pf\"alzische Technische Universit\"at Kaiserslautern-Landau, 67663 Kaiserslautern, Germany}

	\author{Ritwik Mondal}
	\affiliation{Department of Physics, Indian Institute of Technology (Indian School of Mines), 826004 Dhanbad, India}

	\author{Akashdeep Kamra}
	\affiliation{Department of Physics and Research Center OPTIMAS, Rheinland-Pf\"alzische Technische Universit\"at Kaiserslautern-Landau, 67663 Kaiserslautern, Germany}

\begin{abstract}
	The recent proposal and observation of spin inertia, and the consequent high-frequency spin nutation mode, have raised key questions for our understanding of magnetization dynamics, especially considering its high relevance for magnetic memories and ultrafast switching. Notwithstanding recent progress, a clear identification of spin inertia's physical origin thereby offering predictive power remains to be accomplished. Here, discussing general principles for identifying this physical origin, we examine unquenched orbital angular momentum (OAM) finding it to be a key candidate, despite its typically small value. Treating OAM and spin within a two-sublattice model, we derive the equivalent single-sublattice framework for magnetization dynamics making appropriate approximations. The latter naturally manifests the spin inertia term and parameter, which are otherwise introduced phenomenologically. The inertia parameter evaluated within our model is found to be in good agreement with its experimentally observed value in cobalt. We further delineate key experimental signatures that could verify or rule out the unquenched OAM as the origin of the observed high-frequency mode, and avoid a spurious optical mode in a two-sublattice ferromagnet from being identified as nutation. Our analysis offers a potential link between the recently emerged fields of orbitronics and spin inertia, thereby motivating investigations at their intersection.
\end{abstract}

\maketitle


\section{Introduction}

The magnetic statics as well as dynamics in ordered magnets is effectively described by the phenomenological Landau-Lifshitz-Gilbert (LLG) framework \cite{Landau1935, Gilbert2004, Lakshmanan2011} developed as a classical field theory in terms of the position and time-dependent magnetization. Due to the close relation between magnetic moment and (spin) angular momentum \cite{Barnett1935}, the magnetization dynamics and its phenomenology are often understood or motivated by comparing them to the angular momentum dynamics of a spinning top \cite{Gilbert2004}. In order to describe various phenomena and materials, guided by symmetry constraints, phenomenological terms and parameters are introduced which are then determined by comparing with experiments~\cite{Dutta2025, Neeraj2021, Vivek2022}. At the same time, the Landau-Lifshitz framework can be essentially derived as the classical and mean-field limit of the quantum description of magnets \cite{Edwards2009, Wieser2015, Norambuena2020}. Such theories also enable microscopic evaluation of the parameters that enter the LLG framework thereby offering valuable guidance in material selection and hybrid design.

A key sector where LLG phenomenology proves particularly valuable is in describing nanomagnets, used as memory cells, for example \cite{Rana2022, Barman2020, Usadel2006, Tay2013, Biziere2010}. One then makes the so-called macrospin approximation by assuming the magnetization to be spatially homogeneous in the small nanomagnet, on account of the large exchange energy. The simplified description adequately captures the magnetization dynamics including ferromagnetic resonance (FMR) modes \cite{Usadel2006}, the effect of spin transfer torque \cite{Tay2013} as well as magnetic memory switching \cite{Biziere2010}. The time scale for the latter is of special interest and it generally decreases with an increased damping as well as magnetic resonance frequency. Thus, existence of high-frequency magnetization dynamics modes and their understanding is expected to play a crucial role in the performance and design of magnetic memories.   

In a quest to better understand magnetization dynamics by exploiting its analogy with angular momentum dynamics of a spinning top, it has been motivated that the LLG framework should harbor additional ``spin inertia'' terms \cite{Ciornei2011,Titov2021b}. This inclusion is sometimes motivated via an expansion of the damping term in terms of an increasing order of the time derivatives such that the LLG equation takes the form: 
\begin{align}\label{eq:inertia_phenom}
\bm{\dot{\hat{\textbf{m}}}} & = - |\gamma| (\vu{m} \times \mu_0 \va{H}_{\mathrm{eff}}) + \vu{m} \times \left( \alpha \bm{\dot{\hat{\textbf{m}}}} + \kappa \bm{\ddot{\hat{\textbf{m}}}} + \cdots \right),
\end{align}
where $\vu{m}$ is the magnetization unit vector, $\gamma$ is the gyromagnetic ratio, $\va{H}_{\mathrm{eff}}$ is the effective field experienced by the magnetization, and $\alpha$ is the Gilbert damping constant. Here, $\kappa$ becomes the spin inertia parameter, with dimension of time, as it quantifies the term involving the second time derivative, which is typically associated with mass or inertia. The second derivative in Eq.~\eqref{eq:inertia_phenom} gives rise to a second high-frequency resonance mode, besides the standard FMR, which is termed spin nutation making an analogy with the spinning top \cite{Ciornei2011, Wegrowe2012}. Besides these phenomenological analogies, recent attempts to identify the physical origin of spin inertia include consideration of spin-orbit coupling of an electron in vacuum within a Dirac equation framework \cite{Mondal2017}, the classical mechanical description of a current-carrying loop \cite{Giordano2020}, quantum transport frameworks including time-dependent \ak{non-equilibrium Green’s function techniques} \cite{Bhattacharjee2012,Bajpai2019,Osorio2025}, generalized fluctuation–dissipation theorem in the non-Ohmic regime, dynamical Ruderman–Kittel–Kasuya–Yosida
(RKKY) exchange mediated by conduction electrons \cite{Kachkachi2025,Jansen2025}, \ak{and} coupling to dissipative baths in the magnet \cite{Quarenta2024}. Experimental evidence \ak{for spin inertia} was subsequently provided by ultrafast pump–probe and THz excitation studies, which confirmed the presence of nutation in materials such as CoFeB and permalloy \cite{Neeraj2021,Unikandanunni2022,De2025,Saha2026}. However, a clear inconsistency remains between experimentally measured and theoretically predicted values of the inertial parameter $\kappa$. Experiments on permalloy report $\kappa \sim$ 300 fs \cite{Neeraj2021}, with other studies finding values up to 1.6 ps \cite{De2025}. By contrast, \textit{ab initio} calculations predict much smaller values, typically within a few femtoseconds for transition metals such as Fe, Co, and Ni \cite{Thonig2017,Bajaj2024,Juba2019}. \ak{Besides the recent invigoration of interest in investigating the consequences of inertial dynamics,} including auto-oscillations \cite{He2024b,Rodriguez2024,He2024}, thermal agitation \cite{Titov2021b}, inertial spin waves \cite{Mondal2022,Titov2022,Cherkasskii2024,Cherkasskii2021}, and topology of inertial magnons \cite{Ghosh2026}, \ak{a physical understanding of spin inertia's origin} is highly sought to answer several questions including the frequency at which one should expect the spin nutation mode, materials which are better suited to find it, and conclusively identifying a nutation mode by distinguishing it from \ak{a spurious optical magnetization dynamics mode}.

In typical magnetic solids, the magnetic moment is contributed predominantly by the spin of the participating electrons. The OAM contribution arising from the localized electronic wavefunctions is quenched by the crystal field in most materials \cite{Niemeryer2012}. As a result, the observed Land\'e $g$ factor in most magnetic materials is close to 2, indicating a dominance of the spin ($g = 2$) as opposed to orbital ($g = 1$) angular momentum in determining the magnetization \cite{Meyer1961, Chikazumi2009}. Nevertheless, a clear deviation from 2 of the Land\'e $g$ factor value in nearly every magnetic material testifies to the nonzero, though small, contribution of the OAM to magnetization. In the recent years, it has been shown that while OAM is quenched in equilibrium, it can form an effective channel for angular momentum transport in some metals \cite{Bernevig2005, Jo2024, Wang2025, Go2020, Lee2021, Go2021}. This can be seen as analogous to how the spin polarization in normal metals under equilibrium is zero but they act as efficient generators and transmitters of spin currents under appropriate drives. A wide range of effects capitalizing on OAM transport have been proposed and discovered in the recent years, with a healthy scientific debate around the underlying physics. 

In this article, building on these two seemingly unrelated developments, we examine the potential role of unquenched OAM in underlying spin inertia and nutation. While we take inspiration from the aforementioned OAM transport effects, our analysis is based on the well-established understanding that a small, but finite, OAM remains in all magnets after quenching. Identifying that the Russel-Saunders (RS) coupling \cite{Chikazumi2009, BransdenJoachain2003, Liu2011, Condon1952, Bagus2008} acts as an effective exchange interaction between the spin and OAM, we treat the unquenched OAM as an effective second sublattice thereby treating the emerging magnetization dynamics via the two-sublattice model. Under appropriate approximations, applicable in typical materials, we show that this directly leads to the spin inertia term when deriving the spin magnetization dynamics. The consequently evaluated spin inertia parameter is found to be in reasonable agreement with the experimental observation. We further outline some general physical principles that could guide the identification of potential physical origin(s) of spin inertia and nutation.


\section{General principles and summary of results}

In this section, we outline the general principles that we have employed in identifying OAM as the likely origin of spin inertia. Then, we summarize the key results of our work while also presenting the structure of the paper. 

We first employ time-reversal symmetry in establishing that the spin inertia term cannot be rooted in dissipation. In the context of Eq.~\eqref{eq:inertia_phenom}, time reversal leads to the following transformations: $\vu{m} \to - \vu{m}$, $d/dt \to - d/dt$, \ak{and $\va{H}_{\mathrm{eff}} \to - \va{H}_{\mathrm{eff}}$} \cite{Bose2011}. Thus, we see that only the Gilbert damping term ($\propto \alpha$) changes sign under the time-reversal operation due to the natural reason that it captures dissipation or energy loss from the system, which is not recovered under time-reversal. The fact that the spin inertia term ($\propto \kappa$) does not change sign indicates that it is not related to dissipation.

The second principle is that the introduction of a second time derivative, via the spin inertia term, in the LLG equation necessarily increases the order of the secular equation thereby adding a resonance mode. Such an addition cannot be done without increasing the degrees of freedom. Thus, to justify spin inertia, we need a new degree of freedom to couple with the magnetization.

In a solid, there are, in principle, an infinitely many degrees of freedom that could couple to the magnetization. However, focusing on couplings that do not cause dissipation and revisiting the microscopic derivation of magnetization models, we immediately realize that the OAM, although quenched and small, is an omnipresent degree of freedom \cite{Chikazumi2009, Jo2024}. Then, the question becomes why we have been able to ignore it in considering the magnetization dynamics. This partly motivates our study and we find that accounting for the small OAM indeed gives rise to a term and mode that have the same form as spin inertia and nutation, producing results in agreement with expectations. A related question about OAM, without any reference to spin inertia, has been discussed considering an antiferromagnet \cite{Satoh2017}. 

\begin{figure*}[tbh!]
	\centering
		\subfloat[][]{\includegraphics[width=.6\textwidth]{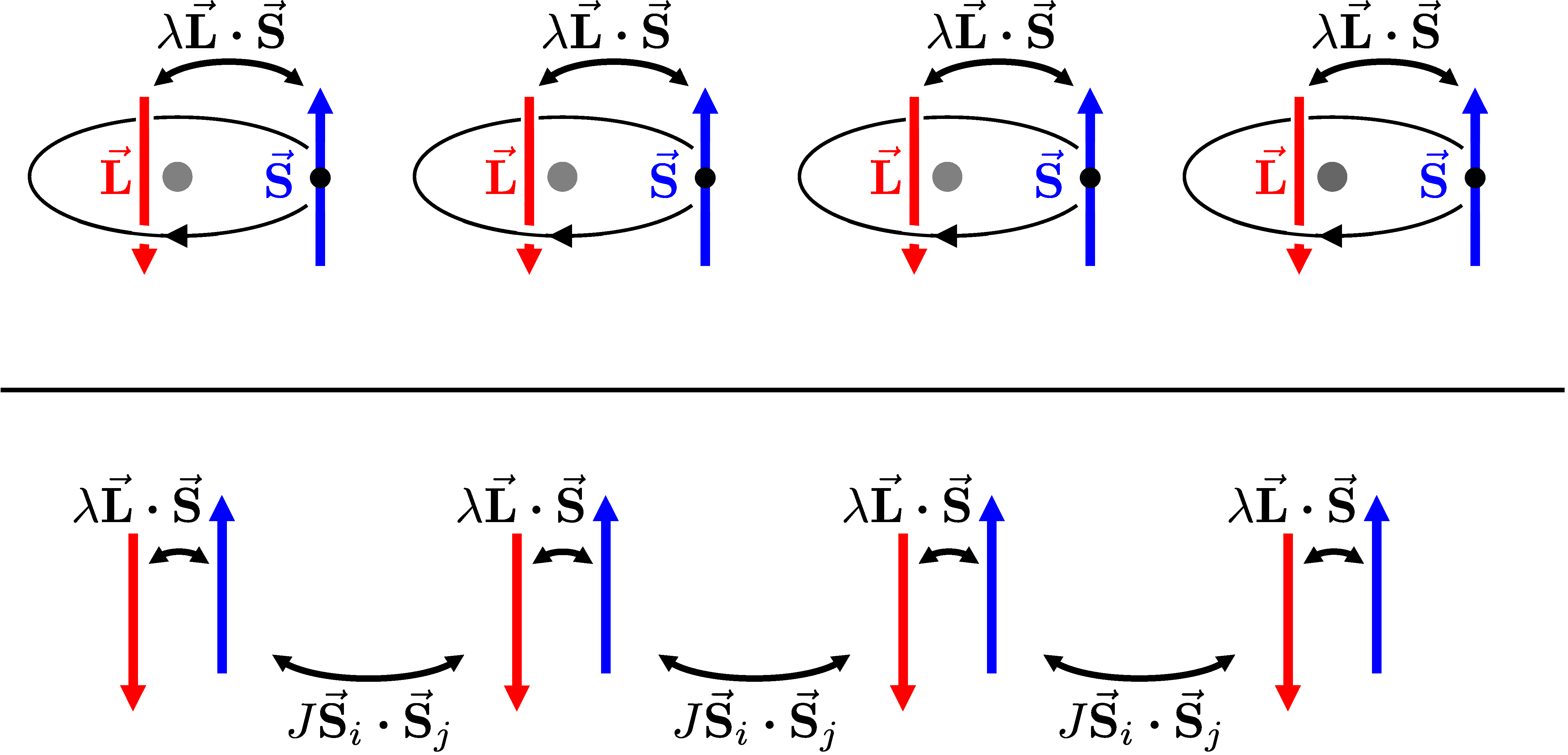}\label{lattice}}
		\hspace{.09\textwidth}
		\subfloat[][]{\includegraphics[width=.25\textwidth]{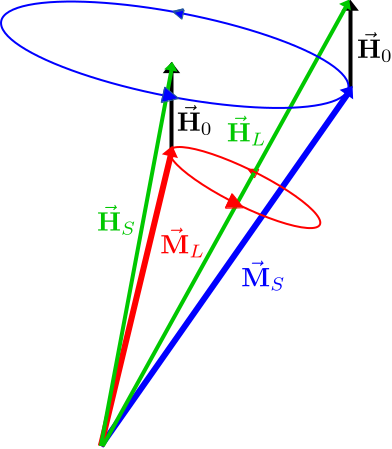}\label{precession}}
	\caption{Schematic depiction of the model. (a) Visualization of Russel-Saunders (RS) coupling. Top panel: An electron in black carrying its spin $\va{S}$ in blue orbiting around the atom core in grey with the orbital angular momentum $\va{L}$. These angular momenta couple on the atomic level via RS coupling with the coupling parameter $\lambda$. This repeats at each atomic site. Bottom panel: The RS coupling is distinct from the exchange coupling $J$ between the spins of electrons at different sites $S_i$ and $S_j$. (b) Visualization of nutation from coupled precessions. The spin magnetization $\va{M}_S$ in blue and the orbital magnetization $\va{M}_L$ in red each precess around their effective magnetic field $\va{H}_S$ for $\va{M}_S$ and $\va{H}_L$ for $\va{M}_L$ in green.}
    \label{fig:schematic}
	\end{figure*}

With this motivation and focusing on ferromagnets, in Sec.~\ref{S2}, we consider a two-sublattice model (Fig.~\ref{fig:schematic}) treating spin and OAM as two different sublattices and contributing magnetizations with the different Land\'e $g$ factors of 2 and 1, respectively. The spin and OAM are assumed to be coupled via a specific form of the atomic spin-orbit interaction - the RS coupling, which mimics exchange interaction and can be ferromagnetic or antiferromagnetic depending on the material \cite{Chikazumi2009}. In considering realistic materials, we assume that the OAM magnitude is much smaller than the spin \cite{Meyer1961, Chikazumi2009}. Evaluating the magnetization dynamics, we find that the OAM degree of freedom indeed gives rise to a high-frequency magnetization dynamics mode, besides the expected FMR, which could be identified as spin nutation. Furthermore, the magnetization precession sense of this nutation mode depends on the sign of the RS coupling. We also find that the magnetization associated with the nutation mode excitation becomes small in the limit of small OAM. This is consistent with the nutation mode being hard to measure due to its reduced coupling to the external drives or environment. The same inference is further corroborated by evaluating the dynamical magnetic susceptibility in Appendix~\ref{susceptibility} and finding it to be small for the nutation mode. This also answers the question posed above from a purely theoretical standpoint: we have been able to ignore the small OAM in studying magnetization dynamics because the additional mode contributed by the presence of OAM hardly couples to external drives.
		
In Sec.~\ref{sec:effectiveLLG}, we derive an effective dynamical equation for the spin magnetization by eliminating the OAM from the coupled two-sublattice dynamics \cite{Kittel1949}. Making the approximation relevant for actual materials, that RS coupling is much stronger than applied fields and that OAM is much smaller than the spin, we obtain an effective equation that contains the spin inertia term in addition to a comparable non-universal contribution that depends on the magnetic anisotropy. Employing the resulting expression for the inertia parameter $\kappa$ in Sec.~\ref{realvalues}, we obtain estimates in reasonable agreement with the experiments conducted on cobalt providing further confidence in our assumed origin of the spin inertia.

Our analysis brings a key point to attention: How should one distinguish an ``optical'' mode in a two or more-sublattice ferromagnet from a spin nutation mode. \ak{This is especially pertinent since many of the magnetic materials, such as the garnets, have two-or more sublattices.} To address this issue, our theoretical analysis in Sec.~\ref{sec:test} suggests measurement of the alleged nutation resonance for different applied magnetic fields. We find that the effective g-factor, obtained via the slope of the mode dispersion with the applied field, is expected to be around 1 for the OAM-based nutation mode. On the other hand, a spurious optical mode stemming from predominantly spin magnetization in a two or more-sublattice ferromagnet will manifest a g-factor close to 2. This same measurement can be used to test whether or not OAM is the origin of the observed high-frequency nutation mode. We conclude with discussion and outlook in Sec.~\ref{sec:discuss}. If OAM is indeed confirmed as the origin of spin inertia, we anticipate effective approaches for controlling it using the recently established knobs of the steady-state OAM content via orbitronic techniques, \ak{such as orbital pumping via magnetization dynamics~\cite{Go2025}}.


\section{Coupled spin and orbital moments as a two-sublattice magnet}
\label{S2}
We know that for every atom, each of its electrons contributes to the magnetic moment with its spin ($s$) and orbital angular momentum (OAM) ($l$)~\cite{Chikazumi2009, Coey2010, Kittel2005}. For lighter atoms, the \ak {Coulombic-origin} $s-s$ and $l-l$ couplings \ak{among} the electrons dominate over the spin-orbit coupling ($l-s$) \cite{Condon1935}. So, all individual electron spin angular momenta of an atom combine to form the total spin vector ($\bm{\vec{S}}$), and all individual orbital angular momenta combine to form the total OAM vector ($\bm{\vec{L}}$). Then an effective coupling between $\bm{\vec{L}}$ and $\bm{\vec{S}}$ is considered which is also known as Russell-Saunders coupling (RS coupling)~\cite{BransdenJoachain2003, Liu2011, Condon1952, Bagus2008}. In the conventional description of magnetism, the OAM is disregarded as it is quenched by the crystal field and expected to be much smaller than the spin. In this section, we take the resulting small OAM explicitly into account treating it as a second magnetic sublattice.

In our model ferromagnetic system, we consider magnetizations due to the total spin and OAM to be ${\va{M_S}}$ and ${\va{M_L}}$ respectively as:
\begin{align*}
   {\va{M}}_S=-\abs{\gamma_S}\frac{1}{V_A}{\va{S}}&=g_S\frac{e^-}{2m_e}\frac{1}{V_A}{\va{S}},\\
   \va{M}_L=-\abs{\gamma_L}\frac{1}{V_A}\va{L}&=g_L\frac{e^-}{2m_e}\frac{1}{V_A}\va{L}.
\end{align*}
 where the $\gamma_\mathcal{L}$ $(\mathcal{L} = L,S)$ are the respective gyromagnetic ratios that depend on the Land\'e $g$-factors $g_S=2$ and $g_L=1$, as well as the negative electron charge $e^-$ and the electron mass $m_e$~\cite{Chikazumi2009, Coey2010, Kittel2005}. $V_A$ is the volume of the primitive unit cell, so that ${1}/{V_A}\ \va{S}$ and ${1}/{V_A}\ \va{L}$ are the material's spin and OAM densities, respectively. 
      
When the outer shell of an atom is less than half-filled, ${\va{L}}$ becomes antiparallel to ${\va{S}}$, giving rise to antiferromagnetic RS coupling. Similarly, for atoms whose outer shell is greater than half-filled, the interaction between ${\va{L}}$ and ${\va{S}}$ becomes ferromagnetic in nature~\cite{BransdenJoachain2003}. To cover both these situations, we write the free energy density due to their interaction as $-\mathfrak{s}_o \lambda\va{M}_s\cdot\va{M}_L$. Here $\lambda$ is the positive RS coupling parameter. $\mathfrak{s}_o$ is used to represent the nature of the coupling with $\mathfrak{s}_F=+1$ for ferromagnetic and $\mathfrak{s}_A=-1$ for antiferromagnetic coupling.

\ak{Before proceeding further, we pause to note the validity, limitations, and potential generalizations of our assumed two-sublattice model for the total spin and OAM at each magnetic ion. Conveniently, direct analogies exist between our assumed two-sublattice model and similar effective treatments of multi-sublattice ferrimagnets, such as YIG~\cite{Princep2017} and GdIG~\cite{Gepraegs2016}. Our model is valid when the Coulombic-origin $s-s$ and $l-l$ exchange couplings are much stronger than all the other energy scales including the relativistic-origin RS coupling~\cite{Chikazumi2009}, frequency of the magnetization mode, and temperature. This assumption is expected to hold in a broad range of materials and conditions. Furthermore, since the magnetic ordering critical temperatures are roughly equal to the weakest inter-ion spin-spin exchange energy, the typically stronger intra-ion spin-spin exchange renders our assumed model a good approximation even close to the ordering temperature. Making a direct analogy with the modeling of multisublattice ferrimagnets via a two-sublattice model~\cite{Princep2017,Gepraegs2016}, our considerations here can be further generalized to capture even higher frequencies comparable to the intra-ion $s-s$ and $l-l$ coupling scales by allowing for more than two-sublattices that account for $s-s$ and $l-l$ couplings explicitly.}


Now, we describe the dynamics of the magnetization unit vectors ($\va{m}_S=\va{M}_S/M_{S0}$ and $\va{m}_L=\va{M}_L/M_{L0}$, with $M_{S0}=|\va{M}_S|$ and $M_{L0}=|\va{M}_L|$) employing  the LLG framework~\cite{Gilbert2004, Kamra2018, Ghosh2024, Dhali2024, Mondal2022, Rozsa2013, Ghosh2025, Mondal2023}. There we take the macrospin approximation, assuming uniformity of the individual magnetization in space, to focus on the zero wavenumber modes. The LLG equations for the magnetization vectors of the two sublattices with intra- and cross-sublattice damping terms are~\cite{Kamra2018, Dhali2024}:    
\begin{align}
    \bm{\dot{\hat{\textbf{m}}}}_S =& -\abs{\gamma_S}\vu{m}_S \cp\mu_0\va{H}_S \nonumber\\&+ \alpha_{SS}\vu{m}_S \cp\bm{\dot{\hat{\textbf{m}}}}_S +\alpha_{SL}\vu{m}_S\cp\bm{\dot{\hat{\textbf{m}}}}_L\label{dglS},\\
    \bm{\dot{\hat{\textbf{{m}}}}}_L =& -\abs{\gamma_L}\vu{m}_L \cp\mu_0\va{H}_L \nonumber \\&+\alpha_{LS}\vu{m}_L\cp\bm{\dot{\hat{\textbf{m}}}}_S + \alpha_{LL} \vu{m}_L \cp\bm{\dot{\hat{\textbf{m}}}}_L\label{dglL},
\end{align}
where, $\mu_0$ is the vacuum permeability, $\alpha_{SS}$, $\alpha_{LL}$ are intra-sublattice Gilbert damping parameters, $\alpha_{SL}$, $\alpha_{LS}$ are cross-sublattice Gilbert damping parameters~\cite{Kamra2018} with ${{\alpha_{SL}}/{\abs{\gamma_S}M_{L0}}={\alpha_{LS}}/{\abs{\gamma_L}M_{S0}}}$. Solving these equations we will be able to calculate the eigen-frequencies and eigenmodes of the system. The effective magnetic fields $\va{H}_{\mathcal{L}}$ $(\mathcal{L} = L,S)$ are calculated from the magnetic free energy $\mathrm{F}$ as its variations with respect to the components of the magnetic moments~\cite{Laxman2011, Ghosh2024, Mondal2022, Rozsa2013, Ghosh2025, Mondal2023, Dhali2024, Evans2014, Skubic2008, Shovon2022, Arpita2024}:
\begin{equation}
    \mu_0H_{\mathcal{L}k}=-\frac{\var\mathrm{F}}{\var M_{\mathcal{L}k}}\ (k = x,y,z)\label{variation}.
\end{equation}
While the framework is general, we consider the following free energy for concreteness~\cite{Kamra2018}:
\begin{align}
    \mathrm{F} &= \int \bigg(-\mu_0\va{H}_0\vdot\qty(\va{M}_S+\va{M}_L)-K_S{{M}_{Sz}}^2-K_L{{M}_{Lz}}^2 \nonumber\\ & \quad \quad-\mathfrak{s}_o \lambda\va{M}_s\vdot\va{M}_L \bigg)\dd{r^3}\label{freeenergy},
\end{align}
where the first term describes the effect of external magnetic field $\bm{\vec{H}}_0$, the second and third terms describe the energy due to easy axis anisotropy with the positive anisotropy parameters $K_\mathcal{L}$ and the last term describes the energy due to the RS coupling between $\bm{\vec{M}}_L$ and $\bm{\vec{M}}_S$.

\subsection{Resonance frequencies}

    To analyze the high frequency mode \cite{Ciornei2011, Boettcher2012, De2025, Olive2012}, as a first test of our model, we calculate the eigen-frequencies of this two-sublattice magnet by solving the LLG equations in Eqs.~(\ref{dglS}) and \eqref{dglL}~\cite{Ghosh2024, Rozsa2013, Mondal2022, Dhali2024, Ghosh2025}. The external magnetic field is assumed to be along the anisotropy axis as $\va{H}_0=\vu{e}_zH_0$. We look at the limit of small excitations from the equilibrium configuration, where both magnetic moments are positioned parallel or antiparallel, depending on the sign of RS coupling, along the $z$-axis. So, the magnetic moment can be approximated as~\cite{Rozsa2013, Mondal2022, Ghosh2025}:
	\begin{equation}
		\va{M}_S = \mqty(M_{Sx}\\ M_{Sy}\\M_{S0})\ ,\ \va{M}_L = \mqty(M_{Lx}\\ M_{Ly}\\\mathfrak{s}_o M_{L0}),\label{smallex}
	\end{equation}
	where, $M_{\mathcal{L}x},M_{\mathcal{L}y}\ll M_{\mathcal{L}0}$. We assume $M_{L0}<M_{S0}$ because the OAM in crystal structures is quenched as a result of the broken rotational symmetry \cite{Chikazumi2009, Niemeryer2012}. As a result, $\va{M}_{L}$ is largely governed by the RS coupling while $\va{M}_{S}$ remains close to parallel to $\va{H}_0$. 

\begin{figure*}[tbh]
	\centering
		\subfloat[][]{\includegraphics[width=.45\textwidth]{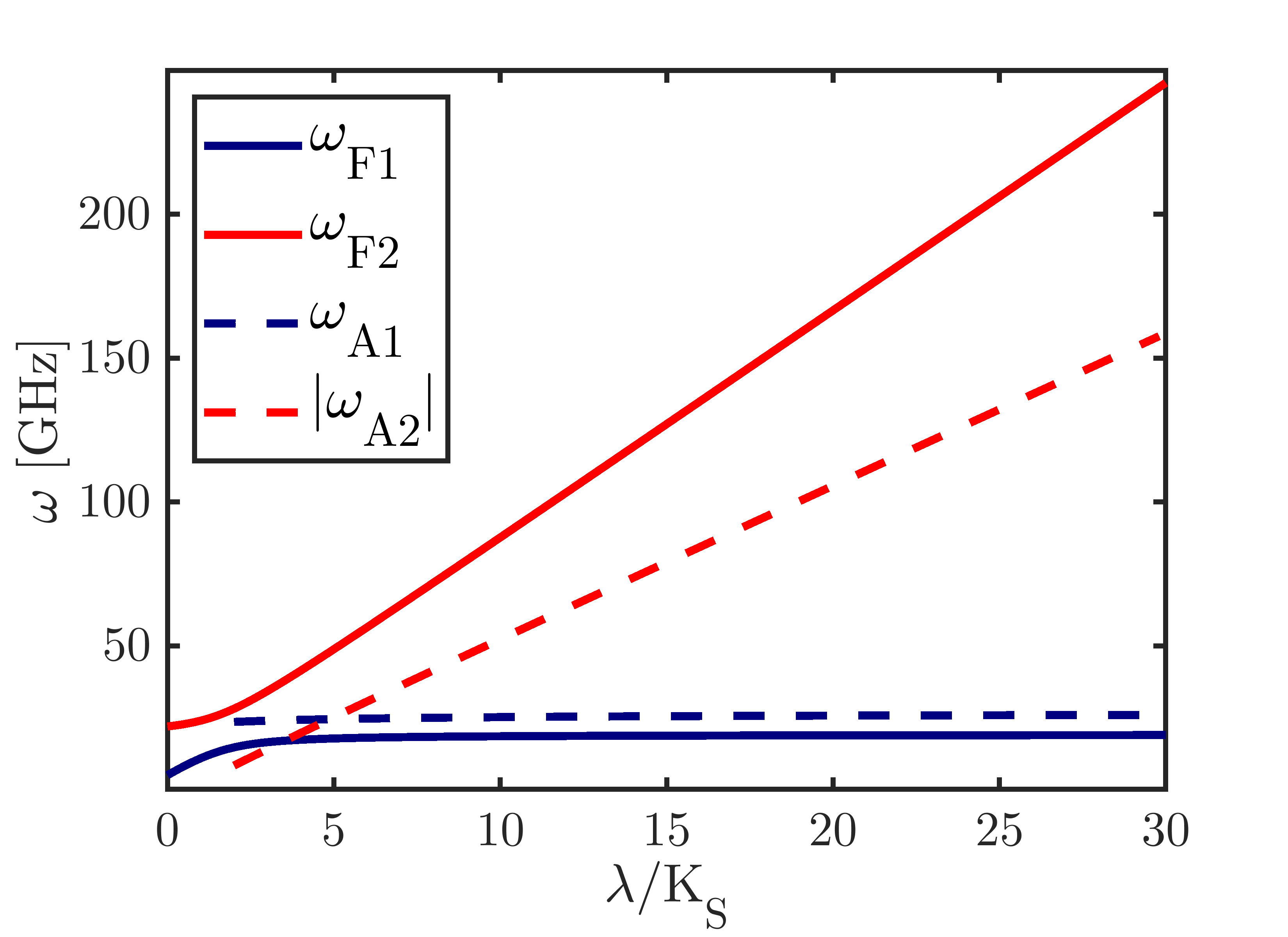}\label{frequencies a}}
		\subfloat[][]{\includegraphics[width=.45\textwidth]{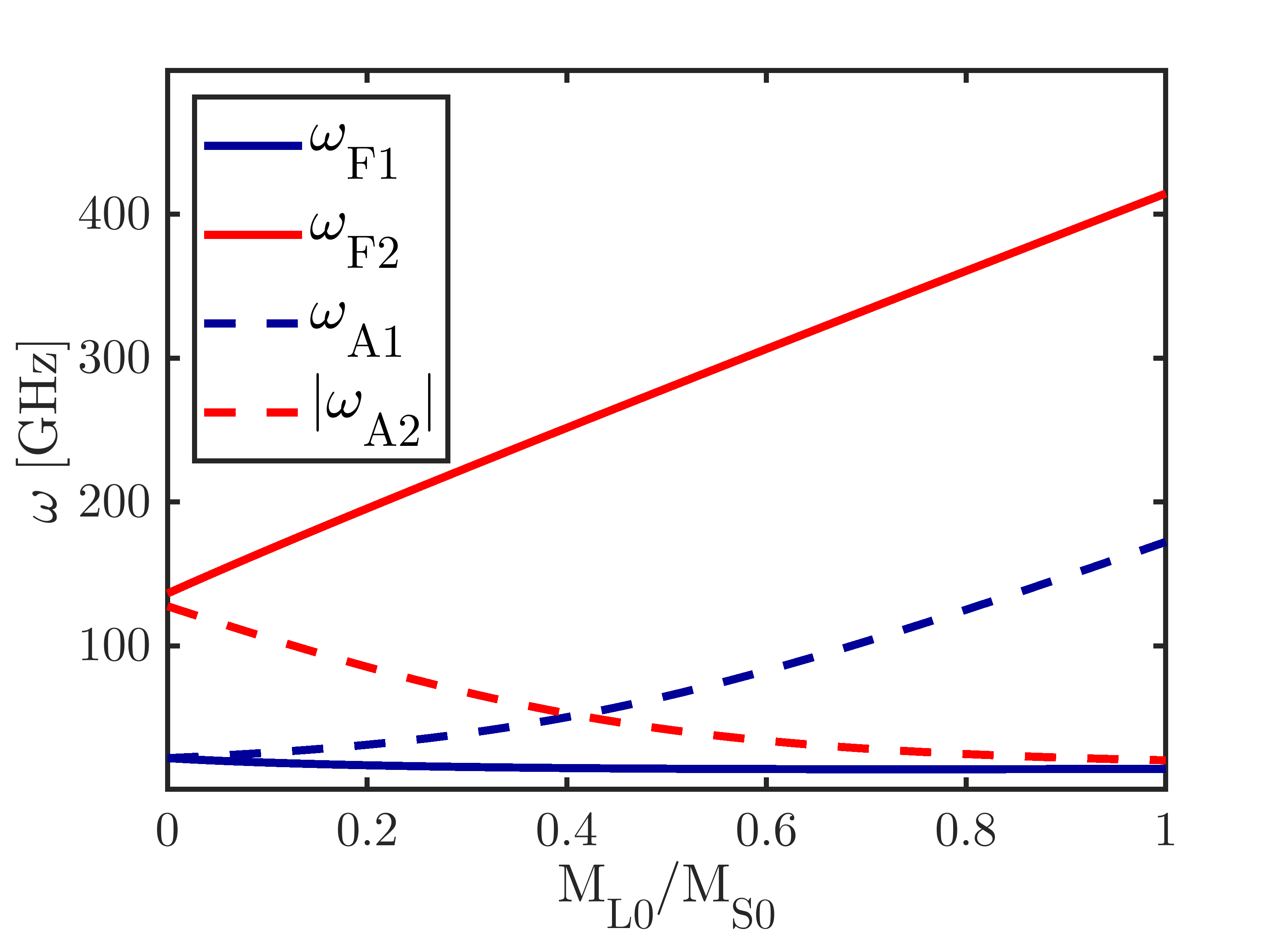}\label{frequencies b}}
		\caption{\label{frequencies} Dependence of the precession and nutation frequencies for ferromagnetic and antiferromagnetic RS coupling on the relative strength of $\lambda$ (a) and $M_{L0}$ (b). For a typical material, $\lambda/K_S$ is estimated at $10^4$ such that the two frequencies are far apart from each other. The parameters are chosen to represent typical material values \cite{Laughin2014, Chikazumi2009} at $M_{S0}=1.5\ \text{J$\cdot$T$^{-1}\cdot$cm$^{-3}$}$, $M_{L0}/M_{S0}=0.1$, $K_S=K_L=2.5\times10^{-2}\ \text{cm$^3\cdot$T$^2\cdot$J}^{-1}$, $\mu_0H_0/K_SM_{S0}=2$, $\abs{\gamma_L}=1.76\times10^{11}\ \text{s$^{-1}\cdot$T}^{-1}$ and $\abs{\gamma_S}/\abs{\gamma_L}=2$ for (a) and $M_{S0}=1.5\ \text{J$\cdot$T$^{-1}\cdot$cm}^{-3}$, $K_S=K_L=2.5\times10^{-2}\ \text{cm$^3\cdot$T$^2\cdot$J}^{-1}$, $\mu_0H_0/K_SM_{S0}=2$, $\lambda/K_S=40$, $\abs{\gamma_L}=1.76\times10^{11}\ \text{s$^{-1}\cdot$T}^{-1}$ and $\abs{\gamma_S}/\abs{\gamma_L}$=2 for (b). For $\omega_{A2}$ the absolute value is depicted to maintain comparability, as a result there appears to be a crossings of $\omega_{A1}$ and $\omega_{A2}$ but this is only a consequence of the visualization and has no physical meaning. }
	\end{figure*}
    
    We can decouple the set of four equations for $M_{Sx}$, $M_{Sy}$, $M_{Lx}$ and $M_{Ly}$ into two sets of two equations by transforming the components of the magnetic moments into the circular basis as ${M_{\mathcal{L}c}=M_{\mathcal{L}x}+\mathfrak{s}_ciM_{\mathcal{L}y}}$, where the index $c$ distinguishes between the right rotating mode with $\mathfrak{s}_r=-1$ and the left rotating mode with $\mathfrak{s}_l=+1$. We then take a Fourier ansatz with $M_{\mathcal{L}c}=\mathcal{M}_{\mathcal{L}c}e^{i\nu t}$ with the mode amplitude $\mathcal{M}_{\mathcal{L}c}$ and the mode frequency $\nu$ to find the solutions. Relegating the detailed mathematics to Appendix~\ref{Afrequencies}, the LLG equations [Eqs.~(\ref{dglS}) and \eqref{dglL}] are simplified to:
    \begin{widetext}
	\begin{align}
		\mqty(\mathfrak{s}_c\nu-\Omega_S-i\nu\alpha_{SS} & \mathfrak{s}_o\abs{\gamma_S}\lambda M_{S0}-i\nu\alpha_{SL}\frac{M_{S0}}{M_{L0}} \\ \abs{\gamma_L}\lambda M_{L0}-\mathfrak{s}_oi\nu\alpha_{LS}\frac{M_{L0}}{M_{S0}} & \mathfrak{s}_c\nu-\mathfrak{s}_o\Omega_L-\mathfrak{s}_oi\nu\alpha_{LL})\mqty(\mathcal{M}_{Sc}\\\mathcal{M}_{Lc})=0,\label{matrix}
	\end{align}
    \end{widetext}
	where we have defined,
	\begin{equation}
		\begin{aligned}
		\Omega_S&=\abs{\gamma_S}\qty(\lambda M_{L0}+2K_SM_{S0}+\mu_0H_0),\\  				
		\Omega_L&=\abs{\gamma_L}\qty(\lambda M_{S0}+2K_LM_{L0}+\mathfrak{s}_o\mu_0H_0).
		\end{aligned}\label{BOmega}
	\end{equation}
To obtain analytically useful results, we assume the Gilbert damping parameters to be small compared to the other parameters. Now, for a non-trivial solution we set the determinant of the matrix in Eq.~(\ref{matrix}) to zero and solve for $\nu$. We expand the solutions up to the linear order in the Gilbert damping or mode damping. As we find more than one solution, they are labeled by $\nu_{on}$ where the index $o$ distinguishes between ferromagnetic and antiferromagnetic RS coupling and $n$ distinguishes between different solutions of Eq.~(\ref{matrix}). These solutions differ by the sign of $\mathfrak{s}_c$ and an additional sign $\mathfrak{s}_a$ ($a = +,-$), which arises for algebraic reasons. With this we can define the four different values of $n$:
    \begin{enumerate}
        \item $n=1$, when $\mathfrak{s}_c=+1$ and $\mathfrak{s}_a=+1$
        \item $n=2$, when $\mathfrak{s}_c=+1$ and $\mathfrak{s}_a=-1$
        \item $n=3$, when $\mathfrak{s}_c=-1$ and $\mathfrak{s}_a=+1$
        \item $n=4$, when $\mathfrak{s}_c=-1$ and $\mathfrak{s}_a=-1$
    \end{enumerate}
    The solutions are complex and we write them as ${\nu_{o n}=\omega_{o n}+iw_{o n}}$. 
    Identifying the real part with the frequency and the imaginary part with the damping of the modes~\cite{Kamra2018}, we evaluate the frequencies for ferromagnetic and antiferromagnetic RS coupling as:
    \begin{widetext}
	\begin{align}
		\omega_{Fn}&=\frac{\mathfrak{s}_c\qty(\Omega_S+\Omega_L)-\mathfrak{s}_a\sqrt{\qty(\Omega_S-\Omega_L)^2+4\lambda^2\abs{\gamma_S}\abs{\gamma_L}M_{S0}M_{L0}}}{2}\label{omegaferro1}\\
		\omega_{An}&=\frac{\mathfrak{s}_c\qty(\Omega_S-\Omega_L)+\mathfrak{s}_a\sqrt{\qty(\Omega_S+\Omega_L)^2-4\lambda^2\abs{\gamma_S}\abs{\gamma_L}M_{S0}M_{L0}}}{2}\label{omegaanti1}
	\end{align}
    \end{widetext}
	We find $\omega_{o 1}=-\omega_{o 4}$ and $\omega_{o 2}=-\omega_{o 3}$. Because of our ansatz, the sign of $\omega$ signifies the sense of complex rotation. Furthermore, a left rotating mode ($M_{\mathcal{L}l}$) and right rotating mode ($M_{\mathcal{L}r}$) with opposite sign of complex rotation but same total frequency add up to one real mode. This mode has clockwise rotation around the $z$-axis if the frequency of $M_{\mathcal{L}l}$ is negative and anticlockwise rotation if it is positive [see Appendix~\ref{rotationsense}]. Therefore, we only consider $\omega_{o 1}$ and $\omega_{o 2}$ as the two unique solutions. The variation of these modes frequencies with the RS coupling strength $\lambda$ and $M_{L0}$ for ferromagnetic and antiferromagnetic cases are plotted in figure \ref{frequencies}. For convenience of discussion, we call the two modes as precession and nutation.

	For the frequencies calculated for ferromagnetic RS coupling in Eq.~\ref{omegaferro1}, we find that both $\omega_{F1}$ and $\omega_{F2}$ are positive [Fig. \ref{frequencies a}], which means that for these modes the vectors $\va{M}_S$ and $\va{M}_L$ rotate counter-clockwise around the $z$-axis [see Appendix~\ref{rotationsense} for the calculation of sense of rotation]. We further find $\omega_{F1}<\omega_{F2}$ [as seen in Fig.~\ref{frequencies}] which causes us to identify $\omega_{F1}$ as the typical FMR precession frequency and $\omega_{F2}$ as the nutation frequency~\cite{Boettcher2012, De2025, Olive2012}. For realistic values of $\lambda$, the precession mode frequency is independent of the RS coupling. This is consistent with the expectation that the FMR does not depend sensitively on OAM.

	
	For frequencies calculated for antiferromagnetic RS coupling in Eq.~(\ref{omegaanti1}), we find that $\omega_{A1}$ is positive and therefore the mode's net magnetization vectors rotate counter-clockwise around the $z$-axis [Fig.~\ref{frequencies}]. However, $\omega_{A2}$ is negative and the mode's net magnetization vectors rotate clockwise [see Appendix~\ref{rotationsense}]. We again find that if we increase $\lambda$, $\omega_{A1}$ remains independent but $\omega_{A2}$ increases [Fig.~\ref{frequencies a}]. Thus, the precession sense of our identified nutation mode depends on the sign of RS coupling. 
    
    If we however vary the ratio ${M_{L0}}/{M_{S0}}$, the magnitude of the frequency increases for the precessional mode and decreases for the nutation mode. Basically, with increasing $|M_L|$, the external field will try to align $\va{M_L}$ along it which will significantly perturb the antiparallel state. So, the precessional frequency ($\omega_{A1}$) is also affected here, which was not the case for the ferromagnetic coupling $\omega_{F1}$. Thus, here the nutation mode behaves differently from the case of ferromagnetic coupling $\omega_{F2}$. For realistic materials, we anticipate ${M_{L0}}/{M_{S0}} \ll 1$.
    

\subsection{Nature of the eigenmodes}

	We now investigate the magnetization dynamics corresponding to these resonance modes. As the orbital magnetization is much smaller than the spin magnetization, we introduce the small parameter $\xi=M_{L0}/M_{S0} \ll 1$ and linearize the frequencies in Eq.~(\ref{omegaferro1}) and~(\ref{omegaanti1}) by expanding them in $\xi$ and only keep terms of the first order. For ferromagnetic and antiferromagnetic couplings, the cases are discussed separately in the following.
    \subsubsection{Ferromagnetic RS coupling}
	For ferromagnetic RS coupling, the linearized frequencies become:
	\begin{align}
		\omega_{F1}=&2\abs{\gamma_S}K_SM_{S0}+\abs{\gamma_S}\mu_0H_0\nonumber\\
			&-\qty(\frac{2\abs{\gamma_S}K_SM_{S0}+\qty(\abs{\gamma_S}-\abs{\gamma_L}\mu_0H_0)}{\frac{\abs{\gamma_L}}{\abs{\gamma_S}}M_{S0}-2\frac{K_S}{\lambda}M_{S0}-\frac{\abs{\gamma_S}-\abs{\gamma_L}}{\lambda\abs{\gamma_S}}\mu_0H_0})M_{L0}\label{linferro1},
    \end{align}
    \begin{align}
		&\omega_{F2}=\lambda\abs{\gamma_L}M_{S0}+\abs{\gamma_L}\mu_0H_0+\bigg(2\abs{\gamma_L}K_L\nonumber\\&+\frac{\lambda^2\abs{\gamma_S}\abs{\gamma_L}M_{S0}}{\lambda\abs{\gamma_L}M_{S0}-2\abs{\gamma_S}K_SM_{S0}-\qty(\abs{\gamma_S}-\abs{\gamma_L})\mu_0H_0}\bigg)M_{L0}\label{linferro2}.
	\end{align}
	They allow us to see that in the limit of small $M_{L0}$, the precession frequency $\omega_{F1}$ is nearly independent of the RS coupling parameter $\lambda$ whereas the nutation frequency $\omega_{F2}$ grows linearly with it [see Fig. \ref{frequencies a}]. Generally, $\omega_{F1}$ appears to be the slightly corrected FMR precession frequency of $\va{M}_S$ with the first two terms being the expected FMR frequency~\cite{Kittel1948,Rezende2020}. On the other hand, $\omega_{F2}$ strongly depends on the RS coupling and is dominated by the term $\lambda\abs{\gamma_L}M_{S0}$. Therefore, $\omega_{F2}$ remains larger than $\omega_{F2}$ even at very small values of $M_{L0}$.

\begin{figure*}[tb]
		\subfloat[][]{\includegraphics[scale=.04]{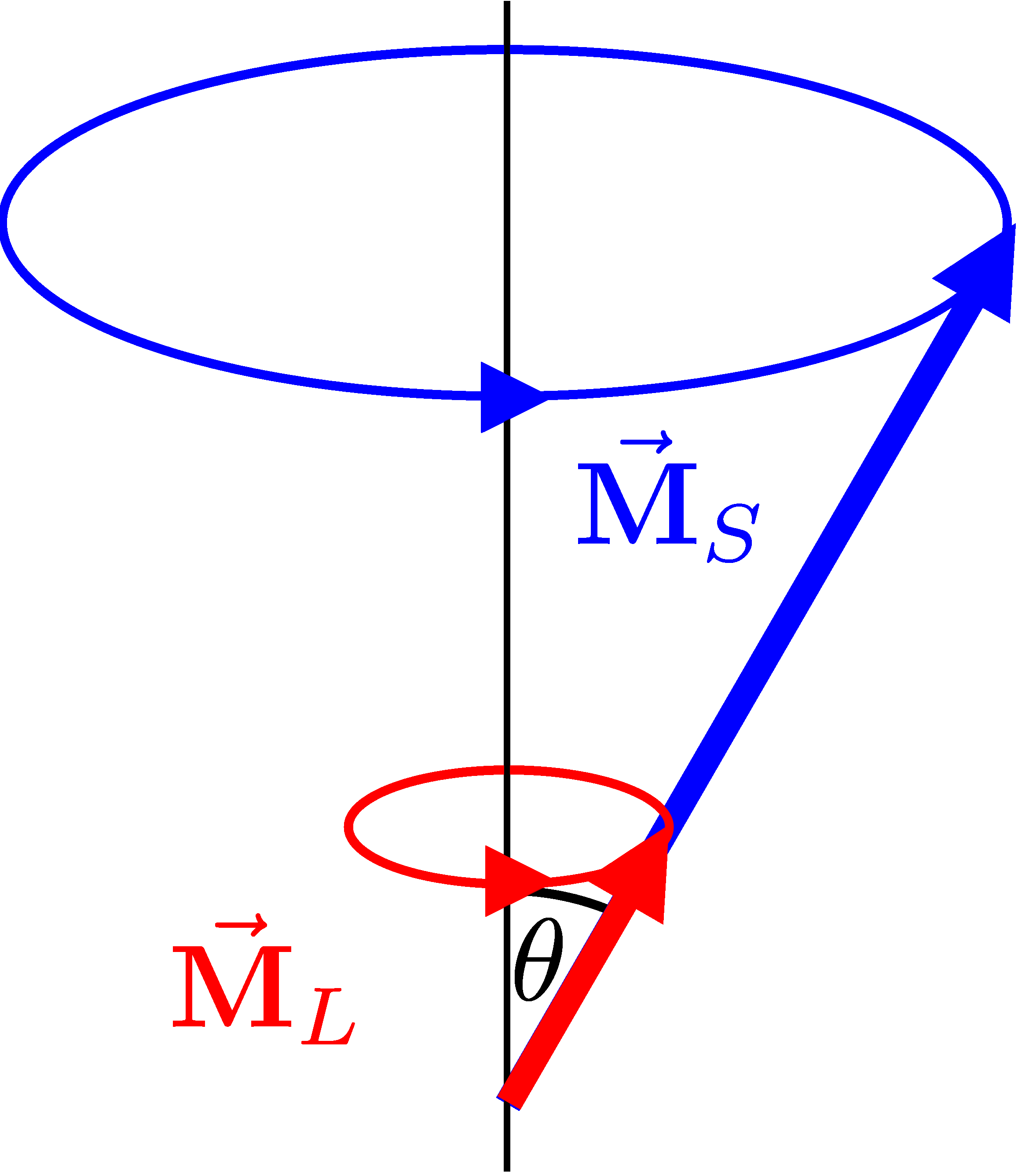}\label{modes a}}
		\hspace{2cm}
		\subfloat[][]{\includegraphics[scale=.04]{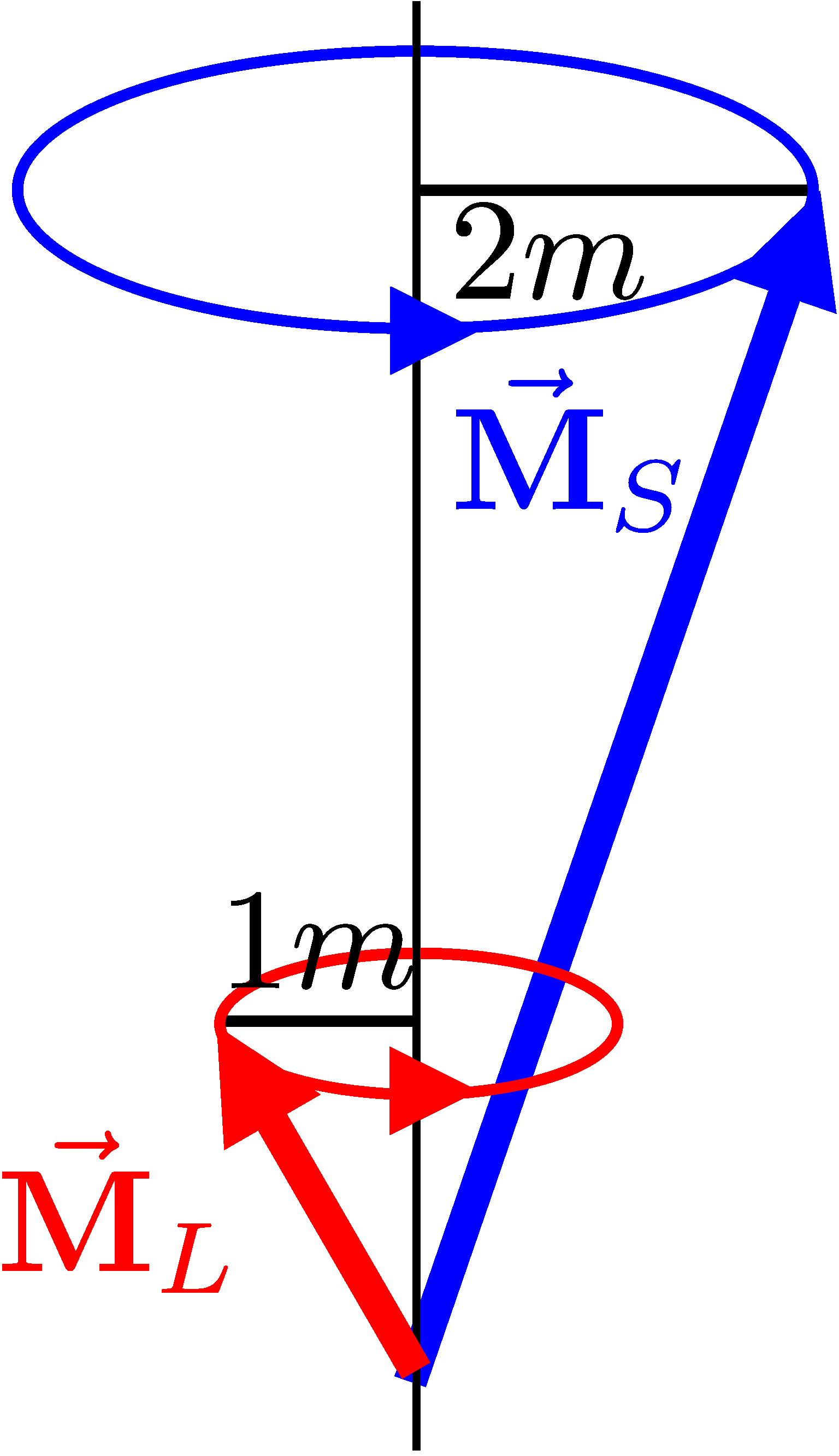}\label{modes b}}\\
		\subfloat[][]{\includegraphics[scale=.04]{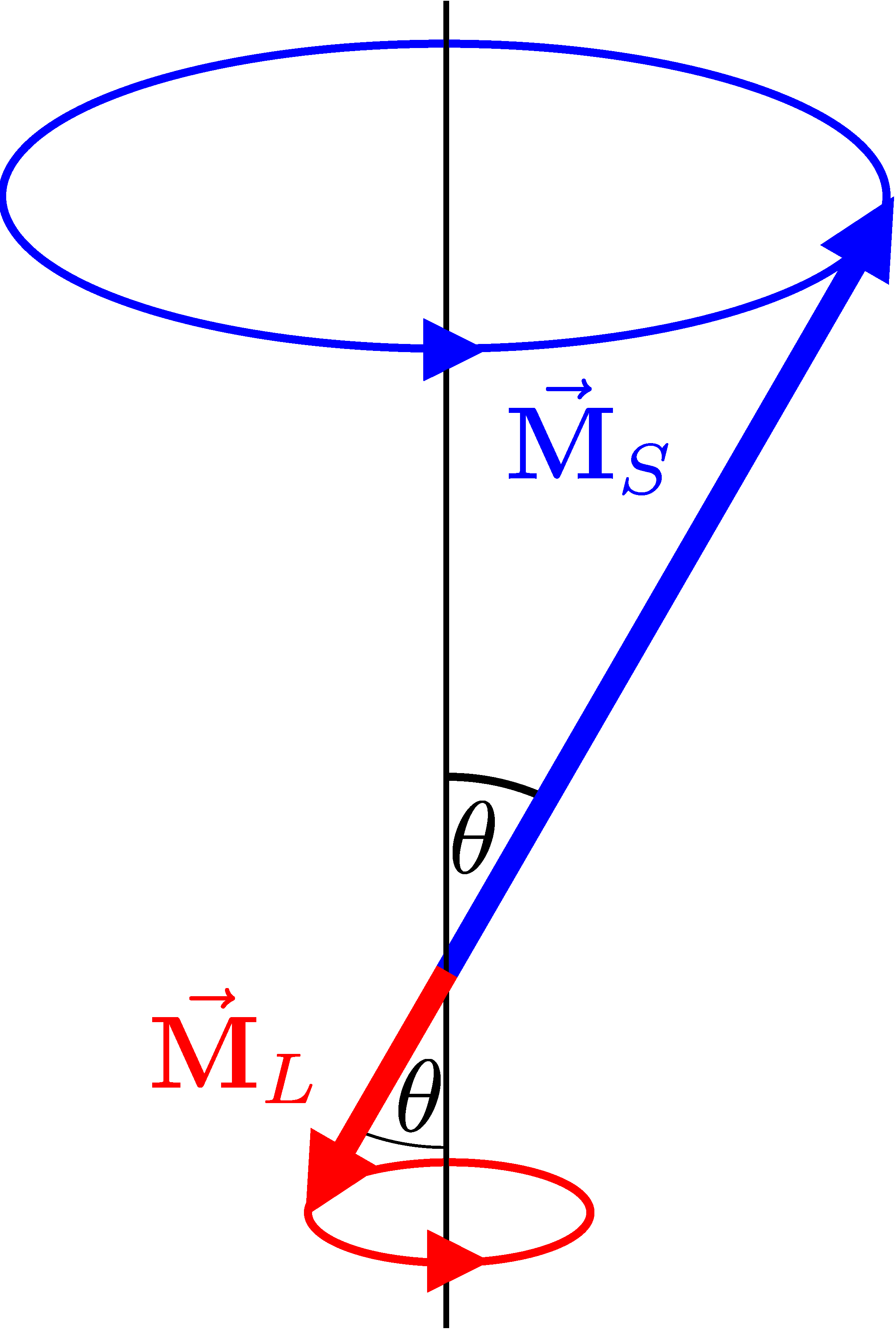}\label{modes c}}
		\hspace{2cm}
		\subfloat[][]{\includegraphics[scale=.04]{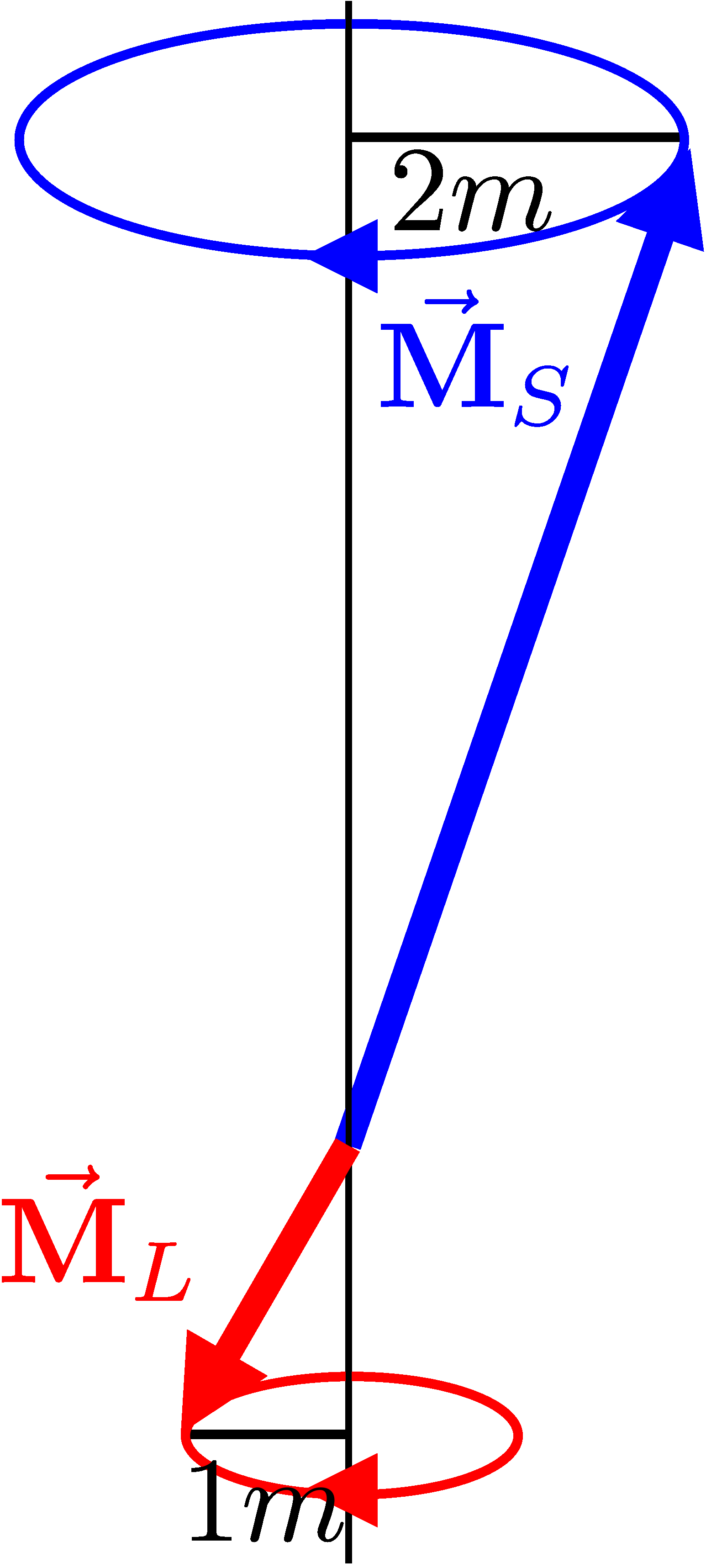}\label{modes d}}
		\caption{\label{vektors}Depiction of the simplified magnetization mode profiles using $\frac{\abs{\gamma_S}}{\abs{\gamma_L}}=2$ and $C_F\approx 1$, $C_A\approx 1$ with (a) the ferromagnetic precession mode with $\omega_{F1}$, (b) the ferromagnetic nutation mode with $\omega_{F2}$, (c) the antiferromagnetic precession mode with $\omega_{A1}$ and (d) the ferromagnetic nutation mode with $\omega_{A2}$. $m$ in (b) and (d) is the magnetic amplitude in $\va{M}_L$.}
	\end{figure*}
    
	 To calculate the mode's magnetization, we insert Eqs. (\ref{linferro1}) and \eqref{linferro2} into Eq.~\eqref{matrix}.
     Thus, we obtain the following relations between the total amplitudes of the spin and OAM ($\mathcal{M}_{SFn}$ and $\mathcal{M}_{LFn}$) of the modes as:
	\begin{equation}
		\mathcal{M}_{SF1}=\frac{1}{C_F\xi}\mathcal{M}_{LF1}\ ,\ 
			\mathcal{M}_{SF2}=-\frac{\abs{\gamma_S}}{\abs{\gamma_L}}C_F\mathcal{M}_{LF2}.
	\end{equation}
    This result for the left rotating circular mode also comes out as the same for the right rotating circular mode. We have defined:
	\begin{align}
		C_F & = \frac{\lambda\abs{\gamma_L}M_{S0}}{\lambda\abs{\gamma_L}M_{S0}-2\abs{\gamma_S}K_SM_{S0}-\qty(\abs{\gamma_S}-\abs{\gamma_L})\mu_0H_0}, \nonumber \\
        & \approx 1.
	\end{align}
    For strong RS coupling, where $\lambda\gg K_S+\mu_0H_0/M_{S0}$, as is suggested by the values of $\lambda$ for single atoms \cite{Koseki2019} [see Sec.~\ref{realvalues} for a detailed discussion], we find $C_F\approx1$. As seen in Fig.~\ref{frequencies a} this approximation is consistent with a large separation between precession and nutation frequency. We also introduce an additional small parameter $\eta$ that quantifies the relation between the excitation of the magnetization modes and the total magnetic moments as $\mathcal{M}_{LFn}=\eta M_{L0}$.
	With this, we can write the single mode excitations in the ferromagnetic system as:
    \begin{widetext}
	\begin{align}
	\begin{split}
		\va M_{SF1}=\mqty(\frac{\abs{\eta}}{C_F}\cos(\abs{\omega_{F1}}t)\\\frac{\abs{\eta}}{C_F}\sin(\abs{\omega_{F1}}t)\\1)M_{S0}\ ,\ \va M_{LF1}=&\mqty(\xi\abs{\eta}\cos(\abs{\omega_{F1}}t)\\\xi\abs{\eta}\sin(\abs{\omega_{F1}}t)\\\xi) M_{S0},
	\end{split}
    \label{M_SF1_M_LF1}\\
	\begin{split}
		\va M_{SF2}=\mqty(C_F\frac{\abs{\gamma_S}}{\abs{\gamma_L}}\xi\abs{\eta}\cos(\abs{\omega_{F2}}t)\\C_F\frac{\abs{\gamma_S}}{\abs{\gamma_L}}\xi\abs{\eta}\sin(\abs{\omega_{F2}}t)\\1)M_{S0}\ ,\ \va M_{LF2}=&\mqty(-\xi\abs{\eta}\cos(\abs{\omega_{F2}}t)\\-\xi\abs{\eta}\sin(\abs{\omega_{F2}}t)\\\xi) M_{S0}.
	\end{split}
    \label{M_SF2_M_LF2}
	\end{align}
    \end{widetext}
A detailed derivation of the above expressions can be found in the Appendix~\ref{rotationsense}. We also calculate the excitation angles ${\theta=\arctan(\sqrt{M_x^2+M_y^2}/M_0)}$ as: 
	\begin{equation}
	\begin{aligned}
		\theta_{SF1}=&\arctan(\frac{\abs{\eta}}{C_F})\ ,\ &\theta_{LF1}=\arctan(\abs{\eta}),\\
		\theta_{SF2}=&\arctan(C_F\frac{\abs{\gamma_S}}{\abs{\gamma_L}}\xi\abs{\eta})\ ,\ &\theta_{LF2}=\arctan(\abs{\eta}).
	\end{aligned}
	\end{equation}
	We can further simplify the mode profiles using $C_F\approx 1$ and ${\abs{\gamma_S}}/{\abs{\gamma_L}}=2$ and have depicted this in Figs.~\ref{modes a} and \ref{modes b}.

	It becomes clear that for the FMR precession mode, the spin and orbital magnetic moments have similar excitation angles and their projections on the $x-y$ plane are always along the same direction [Eq. \eqref{M_SF1_M_LF1}]. This means that for the precession mode the magnetization of spin and orbital angular momenta are locked practically parallel. Because of their similar angles but very different amplitudes we will call their relationship ``angle like''.
    
    For the nutation mode, the excitation amplitudes follow the relation $\abs{\mathcal{M}_{SF2}}\approx 2\abs{\mathcal{M}_{LF2}}$ and their projections on the $x-y$ plane are in opposite directions [Eq. \eqref{M_SF2_M_LF2}]. So, this mode acquires a net amplitude of ${\mathcal{M}_{F2}=\abs{\mathcal{M}_{SF2}}-\abs{\mathcal{M}_{LF2}}\approx \frac{1}{2}\abs{\mathcal{M}_{SF2}}\approx\abs{\mathcal{M}_{LF2}}}$. This also means that the total magnetization amplitude is much smaller for the nutation mode than it is for the precession mode, as it is directly limited by ${\mathcal{M}_{LF2} \ll M_{L0} \ll M_{S0}}$. Because the nutation mode has closely related amplitudes for the sublattices while they are at very different angles, we call this relationship between the sublattices ``amplitude like''. Furthermore, since ${\abs{\gamma_S}}/{\abs{\gamma_L}} \approx 2$, the angular momentum of spin and orbital magnetic moments always cancel each other on the $x$-$y$ plane. This means that the mode has vanishingly small angular momentum. The vanishing angular momentum and unquenched OAM-limited small magnetic amplitude helps to understand why the nutation mode is harder to detect and does not couple well with external drives~\cite{Kittel1948, Polder1949, Farle1998}.

    For the RS coupling considered to be ferromagnetic here, the FMR and nutation modes bear magnetization precession in the same sense. This contrasts with the previous assessment~\cite{Mondal2021}, assuming a positive spin inertia parameter, that nutation and FMR modes precess in opposite senses. This difference is attributed to the sign of the spin inertia parameter being positive for antiferromagnetic RS coupling as we now discuss, and as derived in Sec.~\ref{sec:effectiveLLG} below.
    
	\subsubsection{Antiferromagnetic RS coupling}
	For antiferromagnetic RS coupling, the linearized frequencies are
	\begin{align}
		\omega_{A1}&=2\abs{\gamma_S}K_SM_{S0}+\abs{\gamma_S}\mu_0H_0\nonumber\\&-\qty(\frac{2\abs{\gamma_S}K_SM_{S0}+\qty(\abs{\gamma_S}-\abs{\gamma_L}\mu_0H_0)}{\frac{\abs{\gamma_L}}{\abs{\gamma_S}}M_{S0}+2\frac{K_S}{\lambda}M_{S0}+\frac{\abs{\gamma_S}-\abs{\gamma_L}}{\lambda\abs{\gamma_S}}\mu_0H_0})M_{L0}
        \label{linanti1},
    \end{align}
    \begin{align}
		&\omega_{A2}=-\lambda\abs{\gamma_L} M_{S0}+\abs{\gamma_L}\mu_0H_0+\bigg(-2\abs{\gamma_L}K_L\nonumber\\&+\frac{\lambda^2\abs{\gamma_S}\abs{\gamma_L}M_{S0}}{\abs{\gamma_L}\lambda M_{S0}+2\abs{\gamma_S}K_SM_{S0}+\qty(\abs{\gamma_S}-\abs{\gamma_L})\mu_0H_0}\bigg)M_{L0}\label{linanti2}.
	\end{align}
	These relations differ in some details from that of the ferromagnetic case in Eq.~(\ref{linferro1}) and \eqref{linferro2}, most notably the sign and therefore sense of rotation of the nutation frequency $\omega_{A2}$. On the other hand, their dependence on the magnitude of the RS coupling parameter $\lambda$ (as leading contribution) remains the same as for ferromagnetic case.
    
    Here we calculate the following relations between the magnetizations of a mode:
	\begin{equation}
			\mathcal{M}_{SA1}=-\frac{1}{C_A\xi}\mathcal{M}_{LA1}\ ,\ 
			\mathcal{M}_{SA2}=-\frac{\abs{\gamma_S}}{\abs{\gamma_L}}C_A\mathcal{M}_{LA2},
	\end{equation}
	Similar to the ferromagnetic case, the results are same for the left and right rotating mode. We have defined
	\begin{align}
		C_A & = \frac{\lambda \abs{\gamma_L}M_{S0}}{\lambda \abs{\gamma_L}M_{S0}+2\abs{\gamma_S}K_SM_{S0}+\qty(\abs{\gamma_S}-\abs{\gamma_L})\mu_0H_0}, \nonumber \\
        & \approx 1,
	\end{align}
    where the last approximation is again due to the RS coupling being stronger than anisotropy and typical applied fields.
	We then introduce an additional small parameter $\eta$ which is defined via $\mathcal{M}_{LAn}=\eta M_{L0}$ to calculate the mode profiles as:
    \begin{widetext}
	\begin{align}
		\va M_{SA1}=\mqty(\frac{\abs{\eta}}{C_A}\cos(\abs{\omega_{A1}}t)\\\frac{\abs{\eta}}{C_A}\sin(\abs{\omega_{A1}}t)\\1)M_{S0}\ ,\ \va M_{LA1}=&\mqty(-\xi\abs{\eta}\cos(\abs{\omega_{A1}}t)\\-\xi\abs{\eta}\sin(\abs{\omega_{A1}}t)\\-\xi) M_{S0},\\
		\va M_{SA2}=\mqty(C_A\frac{\abs{\gamma_S}}{\abs{\gamma_L}}\xi\abs{\eta}\cos(\abs{\omega_{A2}}t)\\-C_A\frac{\abs{\gamma_S}}{\abs{\gamma_L}}\xi\abs{\eta}\sin(\abs{\omega_{A2}}t)\\1)M_{S0}\ ,\ \va M_{LA2}=&\mqty(-\xi\abs{\eta}\cos(\abs{\omega_{A2}}t)\\\xi\abs{\eta}\sin(\abs{\omega_{A2}}t)\\-\xi) M_{S0},
	\end{align}
    \end{widetext}
	and excitation angles:
	\begin{equation}
	\begin{aligned}
		\theta_{SA1}=&\arctan(\frac{\abs{\eta}}{C_A})\ ,\ &\theta_{LA1}=\arctan(\abs{\eta}),\\
		\theta_{SA2}=&\arctan(C_A\frac{\abs{\gamma_S}}{\abs{\gamma_L}}\xi\abs{\eta})\ ,\ &\theta_{LA2}=\arctan(\abs{\eta}).
	\end{aligned}
	\end{equation}
	The magnetization profiles using $C_A\approx1$ and $\abs{\gamma_S}/\abs{\gamma_L}=2$ are shown in Fig.~\ref{modes c} and \ref{modes d}.
    
    Similarly to the ferromagnetic case, in the FMR precession mode, we find that the spin and orbital magnetizations are locked antiparallel with the same excitation angles, making their relationship angle like. The amplitudes of the dynamical components for the nutation mode follow the relation $\mathcal{M}_{SA2}\approx -2\mathcal{M}_{LA2}$, making the relationship amplitude like. Accordingly, we again find that the nutation mode caries vanishing angular momentum and its magnetization amplitude is small and limited by the smallness of $M_{L0}$. However, for the considered case of antiferromagnetic RS coupling, the nutation and FMR modes precess in the opposite senses, consistent with previous findings~\cite{Mondal2021}.
	
	With this, we conclude our discussion of the magnetization modes for both ferromagnetic and antiferromagnetic RS coupling including frequencies and mode profiles. Despite few obvious differences and notably the opposite precession sense of the nutation mode between the ferromagnetic and antiferromagnetic RS coupling cases, the two cases behave quite similarly. The smaller FMR frequency is largely independent of the strength of the RS coupling, and the magnetizations of the precession mode have an angle like relationship. This can be seen as the recovery of typical magnetization dynamics that disregards OAM altogether. The larger nutation frequency on the other hand is dominated by the RS coupling. The mode's magnetizations have an amplitude like relationship that limits their maximum amplitude due to the smallness of the unquenched OAM and suppresses their coupling to the external world.


\section{Deriving spin inertia term}\label{sec:effectiveLLG}

    Our two coupled first order differential equations for $\va{M}_S$ and $\va{M}_L$ should reduce to a single second order differential equation on elimination of $\va{M}_L$ \cite{Gomonay2010}. Such a reduced equation is expected to be identical to the usual LLG equation, which has been amended with the phenomenological spin inertia term bearing a second order time derivative \cite{Mondal2021, Vivek2022, Neeraj2021, Mondal2023, Ghosh2024, Titov2021, Mondal2017, Cherkasskii2022}. Since, our two-sublattice model employs transparent microscopic physical variables, it should enable a microscopic evaluation of the nutation parameter $\kappa$. With this motivation, we now derive the dynamical equation for $\va{M}_S$ by eliminating $\va{M}_L$~\cite{Kittel1949, Gilbert2004} and making the approximations - RS coupling is large and OAM is small - relevant for typical materials.

	Neglecting Gilbert damping, we obtain from Eqs.~(\ref{dglS}) and \eqref{dglL}:
    \begin{align}
		\dot{\va{M}}_S&=-\abs{\gamma_S}\va{M}_S\cp\mu_0\va{H}_0\nonumber\\&-\abs{\gamma_S}2K_S\va{M}_S\cp{M}_{Sz}\vu{e}_z-\mathfrak{s}_o\abs{\gamma_S}\lambda\va{M}_S\cp\va{M}_L\label{sdot},\\
		\dot{\va{M}}_L&=-\abs{\gamma_L}\va{M}_L\cp\mu_0\va{H}_0\nonumber\\&-\abs{\gamma_L}2K_L\va{M}_S\cp{M}_{Lz}\vu{e}_z-\mathfrak{s}_o\abs{\gamma_L}\lambda\va{M}_L\cp\va{M}_S.\label{ldot}
	\end{align}
    First, we calculate $\va{M}_S\cp\ddot{\va{M}}_S$ from Eq.~\eqref{sdot} and insert Eq.~(\ref{ldot}) into the result. We can then solve for $\va{M}_S\cp\va{M}_L$, insert this result into Eq.~\eqref{sdot} and solve for $\dot{\va{M}}_S$ to obtain the exact solution (see Appendix~\ref{fullcalc} for details):
    \begin{widetext}
	\begin{align}
    &\dot{\va{M}}_S=-\abs{\gamma_L}\abs{\gamma_S}\va{M}_S\cp\Bigg(\frac{\qty(\va{M}_S\vdot\mu_0\va{H}_L+\mathfrak{s}_o\lambda\va{M}_S\vdot\va{M}_L)\mu_0\va{H}_0}{\abs{\gamma_L}\va{M}_S\vdot\mu_0\va{H}_L+\abs{\gamma_S}\va{M}_S\vdot\mu_0\va{H}_S}
    +\frac{2K_S\qty(\va{M}_S\vdot\mu_0\va{H}_L){M}_{Sz}\vu{e}_z}{\abs{\gamma_L}\va{M}_S\vdot\mu_0\va{H}_L+\abs{\gamma_S}\va{M}_S\vdot\mu_0\va{H}_S}\nonumber\\
    &+\frac{\mathfrak{s}_o2\lambda K_L\qty(\va{M}_S\vdot\va{M}_L){M}_{Lz}\vu{e}_z}{\abs{\gamma_L}\va{M}_S\vdot\mu_0\va{H}_L+\abs{\gamma_S}\va{M}_S\vdot\mu_0\va{H}_S}\Bigg)
    -\frac{\va{M}_S\cp\ddot{\va{M}}_S}{\abs{\gamma_L}\va{M}_S\vdot\mu_0\va{H}_L+\abs{\gamma_S}\va{M}_S\vdot\mu_0\va{H}_S}
    -\frac{2K_S\abs{\gamma_S}\va{M}_S\cp\qty(\va{M}_S\cp\dot{M}_{Sz}\vu{e}_z)}{\abs{\gamma_L}\va{M}_S\vdot\mu_0\va{H}_L+\abs{\gamma_S}\va{M}_S\vdot\mu_0\va{H}_S}.
    \label{exact}
	\end{align}
    \end{widetext}
    Here we employed:
    \begin{align}
       \mu_0\va{H}_S=\mathfrak{s}_o\lambda\va{M}_{L}+\mu_0\va{H}_0+2K_S{M}_{Sz}\vu{e}_z \label{H_S},\\
       \mu_0\va{H}_L=\mathfrak{s}_o\lambda\va{M}_{S}+\mu_0\va{H}_0+2K_L{M}_{Lz}\vu{e}_z \label{H_L},
    \end{align}
	as the effective magnetic fields experienced by spin and orbital magnetic moments, respectively.

    We can then use that $\lambda$ is much larger than any other factors to assume that the RS coupling locks $\va{M}_S$ and $\va{M}_L$ near parallel or antiparallel (see Appendix \ref{lock}) and expand all terms in orders of $\lambda$, only keeping the highest. We will further use $M_{S0} \gg M_{L0}$ to only consider terms up to the first order in $M_{L0}/M_{S0}$. Making these approximations and retaining terms up to the linear order in the dynamical components of the magnetization, we obtain the dynamical equation:
    \begin{equation}
		\begin{aligned}
		\dot{\va{M}}_S\approx&-\abs{\gamma_S'}\va{M}_S\cp\mu_0\va{H}_{\text{eff}}+ \frac{\kappa}{M_{S0}}\va{M}_S\cp\ddot{\va{M}}_S\\& + 2\abs{\gamma_{S}}K_S\frac{\kappa}{M_{S0}}\va{M}_S\cp\qty(\va{M}_S\cp \dot{M}_{Sz}\vu{e}_z)\label{finalsingle},
		\end{aligned}
	\end{equation}
    where
	\begin{align}
		\kappa \equiv &-\frac{\mathfrak{s}_o M_{S0}}{\lambda M_{S0}\qty(\abs{\gamma_L}M_{S0}+\mathfrak{s}_o\abs{\gamma_S}M_{L0})},
		\label{inertia}\\
		\abs{\gamma_S'} \equiv &\frac{\abs{\gamma_L}\qty(M_{S0}+\mathfrak{s}_oM_{L0})}{\abs{\gamma_L}M_{S0}+\mathfrak{s}_o\abs{\gamma_S}M_{L0}}\abs{\gamma_S},\\
		K_S' \equiv &\frac{{M}_{S0}}{{M}_{S0}+\mathfrak{s}_oM_{L0}}K_S, \\
        \mu_0\va{H}_{\text{eff}} \equiv & \mu_0\va{H}_0+2K_S'{M}_{Sz}\vu{e}_z
        \label{H_eff_final}.
	\end{align}
    In the equations above, $\kappa$ becomes the nutation parameter and $\gamma_S'$ and $K_S'$ are the new total gyromagnetic moment and anisotropy constant respectively (For more detailed expressions, see Appendix~\ref{lambda}). The new gyromagnetic moment $\gamma_S'$ obtained above agrees with its expected value from Einstein-de-Haas experiments~\cite{Chikazumi2009, Kittel1949}. The dynamical equation ~\eqref{finalsingle} derived above essentially matches the phenomenologically proposed equation including spin inertia \cite{Mondal2021, Vivek2022, Neeraj2021, Mondal2023, Ghosh2024, Titov2021, Mondal2017, Cherkasskii2022}
	\begin{equation}
	\dv{\va{M}}{t}=-\abs{\gamma}\va{M}\cp\mu_0\va{H}_{\text{eff}}+\frac{\kappa}{M_{S0}}\va{M}\cp\dv[2]{\va{M}}{t}\label{goal}.
	\end{equation}
     It only differs in the term $2 \abs{\gamma_{S}}K_S\kappa/M_{S0}\ \va{M}_S\cp(\va{M}_S\cp \dot{M}_{Sz}\vu{e}_z)$, which depends on our assumed anisotropy and is, thus, non-universal.

     Thus, we obtain the approximate and linearized dynamical equation \eqref{finalsingle} for the spin magnetization with the effective parameters expressed in terms of microscopic quantities via Eqs.~\eqref{inertia}-\eqref{H_eff_final}. Our analysis recovers the universal spin-inertia term, introduced phenomenologically in previous studies, and suggests the existence of an additional nonuniversal term that depends on the anisotropy landscape. Furthermore, the sign of our evaluated spin inertia parameter depends directly on the sign of RS coupling, which further correlates with the precession sense of the nutation mode discussed in Sec.~\ref{S2} above offering a key experimental test and observable~\cite{Mondal2021}. Specifically, the spin inertia parameter $\kappa$ is found to be positive for antiferromagnetic RS coupling. 
     
     \ak{Finally, the dependence of the spin inertia parameter $\kappa$ on the net OAM magnetization $M_{L0}$ as per Eq.~\eqref{inertia} uncovers a way to control spin inertial via $M_{L0}$ in a nonequilibrium steady state. This could be achieved using various means including orbital pumping via GHz range magnetization dynamics excitation~\cite{Go2025} which would then directly control spin inertia.}
	
	
\section{Estimating the inertia constant}
    \label{realvalues}
	To estimate values for the nutation frequency and constant, we limit ourselves to the leading contribution, which is the RS coupling. We can write the results from Eq.~\eqref{linferro2} and \eqref{linanti2} unified as:
	\begin{align}
    \omega_{o 2}&=\mathfrak{s}_o\lambda\qty(\abs{\gamma_L}M_{S0}+\mathfrak{s}_o\abs{\gamma_S}M_{L0})\label{nutation_simplified},
	\end{align}
    while $\kappa$ can be found in Eq.~\eqref{inertia}. Assuming $M_{L0} \ll M_{S0}$, we can approximate $M_{S0}$ as equal to $M_0$, the total saturation magnetization. For a more precise calculation $M_{L0}$ can be calculated from the the different gyromagnetic ratios and $g$-factors i.e. $g$ from FMR measurements~\cite{Frait1965,Frait1971} and $g'$ from measurements of the Einstein-de-Haas effect~\cite{Chikazumi2009,Scott1962, Wallis2006}. These $g$-factors are calculated as:
	\begin{equation}
			g=\frac{M_{S0}+\mathfrak{s}_oM_{L0}}{\frac{1}{2}M_{S0}}\ ,\ g'=\frac{M_{S0}+\mathfrak{s}_oM_{L0}}{\frac{1}{2}M_{S0}+\mathfrak{s}_oM_{L0}},
	\end{equation} 
	which allows us to obtain:
	\begin{equation}
		M_{L0}=\mathfrak{s}_o\frac{1}{2}M_{S0}\qty(\frac{g}{g'}-1).
	\end{equation}
	With these, we can calculate the nutation frequencies as:
	\begin{equation}
		\begin{aligned}
			\omega_{o 2}=&\mathfrak{s}_o\frac{1}{2}\lambda M_{S0}\abs{\gamma_S}\frac{g}{g'}.
		\end{aligned}
	\end{equation}
	Alternatively, $M_{L0}$ can be estimated from just one of the Landé $g$-factor as $M_{L0}\approx  \mathfrak{s}_oM_{S0}\qty({g}/{2}-1)$ or $M_{L0}\approx \mathfrak{s}_oM_{S0}\qty(1-{g'}/{2})$ \cite{Chikazumi2009}.
	
	If we use RS coupling values of the single cobalt atoms ($7.7454\times10^{-21} \text{ J}$) \cite{Koseki2019}, cobalt's atomic mass ($58.9\text{ u}$) \cite{Prohaska2022}, values of $g=2.21$ and $g'=1.85$ for cobalt metal \cite{Chikazumi2009}, the density of cobalt ($8.834\text{ g/cm}^3$) \cite{Arblaster2018} to calculate $\lambda=139.3\text{ cm$^3$T$^2$/J}$ and cobalt's saturation magnetization ($1.45\text{ J/T/cm}^3$) \cite{Laughin2014}, the resulting nutation frequency as per our OAM-based model becomes $\omega_{\text{Co} 2}\approx 38.7\times10^{12}\ \text{s}^{-1}$, which is only slightly bigger than the experimental values ($8.2\times10^{12}\text{ s}^{-1}$, $8.8\times10^{12}\text{ s}^{-1}$ and $13.1\times10^{12}\text{ s}^{-1}$)  that were measured as nutation resonances in thin cobalt films \cite{Vivek2022}. The minor discrepancy here is not unexpected, as the strength of the RS coupling differs between free atoms and atoms in crystals \cite{buenemann2008,Goodenough1968} and also depends on the exact electron configuration \cite{Koseki2019}. 
    
    If we instead use values of $\lambda$ that were experimentally obtained for Co$^{2+}$ in CoO ($\lambda=47.1\text{ cm$^3$T$^2$/J}$) \cite{Satoh2017} for our calculation in cobalt metal, we find $\omega_{\text{Co}}=13.1\times10^{12}\ \text{s}^{-1}$, which agrees better with the range of the experimental results from $8.2\times10^{12}\text{ s}^{-1}$ to $13.1\times10^{12}\text{ s}^{-1}$. Using the same values, we obtain $\kappa_{\text{Co}}\approx 76.3\text{ fs}$, which again shows reasonable agreement with the experimental values of $75\text{ fs}$, $110\text{ fs}$ and $120\text{ fs}$ \cite{Vivek2022}. Thus, our OAM-based analysis of spin inertia and nutation seems to be effective for cobalt as a case study.


\section{Distinguishing nutation mode from optical mode in a two-sublattice ferromagnet}\label{sec:test}

High-frequency resonances, identified as spin nutation, have been measured in composite ferromagnets such as permalloy~\cite{De2025}, which comprises of nickel as well as iron. \ak{Such a ferromagnet} will have two or more atoms in its unit cell making it a two or more-sublattice ferromagnet. \ak{Furthermore, many of the widely used magnets, such as the garnets and materials commonly used as permanent magnets, bear more than one sublattices. Thus, these necessarily harbor high-frequency optical magnonic modes.} This assessment is based on the general principle of counting the degrees of freedom. Thus, it is inevitable that magnetic materials harbor high-frequency modes which could be misidentified as nutation. Employing the wider applicability of our two-sublattice model, we now examine this question and delineate a way to identify the spin nutation mode.

As a simple case study, we compare our model with a two-sublattice ferromagnet harboring equivalent sublattices. The latter case is signified by labeling the two sublattices as A and B while the intersublattice ferromagnetic exchange is denotes by replacing the $\lambda$ with $J$. With these replacements, our considered nutation and a spurious optical modes can still be evaluated using Eq.~(\ref{omegaferro1}). On the one hand, we may distinguish them based on the quantitative difference in their frequencies. In most materials, one expects the intersublattice exchange $J$ to be much larger than the RS coupling which translates to a much higher frequency for an optical mode as compared to a nutation mode. However, this approach requires knowledge of these exchange and RS coupling strengths to be sure.

As a better alternative, we propose to examine the dependence of the mode frequencies with the applied magnetic field. In the limit of $M_{L0} \ll M_{S0}$ and using Eq.~\eqref{linferro2}, we obtain the nutation mode frequency as 
\begin{align}
\omega_{F2} = & \lambda\abs{\gamma_L}M_{S0}+\abs{\gamma_L}\mu_0H_0 + \bigg(2\abs{\gamma_L}K_L\nonumber\\
               & +\frac{\lambda^2\abs{\gamma_S}\abs{\gamma_L}M_{S0}}{\lambda\abs{\gamma_L}M_{S0}-2\abs{\gamma_S}K_SM_{S0}-\qty(\abs{\gamma_S}-\abs{\gamma_L})\mu_0H_0}\bigg)M_{L0}, \\
              \approx & \lambda\abs{\gamma_L}M_{S0} + \abs{\gamma_L}\mu_0H_0,
\end{align}
whence we see that the dispersion of the mode frequency $\partial \omega_{F2} / (\mu_0 \partial H_0) $ takes the value $\abs{\gamma_L}$ signifying the OAM origin of this nutation mode. On the other hand, the optical mode in a two-sublattice ferromagnet is obtained in the limit $\Omega_A - \Omega_B \approx 0$ and $J \gg K_{A,B}$ as:
\begin{align}
		\omega_{F2} \approx & \frac{\Omega_A+\Omega_B}{2}+J\sqrt{\abs{\gamma_A}\abs{\gamma_B}M_{A0}M_{B0}}\nonumber\\
         & +\frac{1}{4J\sqrt{\abs{\gamma_A}\abs{\gamma_B}M_{A0}M_{B0}}}\frac{\qty(\Omega_A-\Omega_B)^2}{2}, \\
         \approx & \frac{J\qty(\abs{\gamma_A}M_{B0}+\abs{\gamma_B}M_{A0})}{2} + J\sqrt{\abs{\gamma_A}\abs{\gamma_B}M_{A0}M_{B0}} \nonumber \\
        & + \abs{\gamma_A}K_AM_{A0}+\abs{\gamma_B}K_BM_{B0}+\frac{\qty(\abs{\gamma_A}+\abs{\gamma_B})\mu_0H_0}{2},        \label{omega_Fn_similar_sublattice}
\end{align}
which shows that the mode dispersion is now $\partial \omega_{F2} / (\mu_0 \partial H_0) = (|\gamma_A| + |\gamma_B|)/2 \approx |\gamma_S|$, considering that spin remains the dominant contributor to the magnetization.

Thus, we see that if the nutation mode is indeed caused by OAM, we should find the corresponding effective g-factor from the magnetic field dependence of the mode frequency. On the other hand, the effective g-factor will correspond to that of spin if the mode is an optical mode in a symmetric two-sublattice magnet. The case of optical modes in a multi-sublattice magnet or an asymmetric two-sublattice magnet should be evaluated separately when considering a specific material of interest. Despite potentially different details, our suggestion of examining the mode's magnetic field dependence promises to offer valuable insights into its physical origin.


\ak{\section{Discussion and Conclusion}} \label{sec:discuss}
In summary, we have examined a natural possibility that the unquenched orbital angular momentum (OAM) leads to an additional mode in the magnetization dynamics which could be identified with spin nutation and described via the spin inertia term introduced phenomenologically in previous works. This analysis allowed us to obtain an expression for the spin inertia parameter which was evaluated to be in reasonable agreement with the experimental observations of the spin nutation mode in cobalt. We further employed generic principles, such as counting degrees of freedom and resonance modes, in outlining strategies for avoiding misidentification of optical modes as spin nutation. Furthermore, we propose experimental signatures to test orbital angular momentum as the origin of spin inertia and nutation. 

\ak{Specifically, we find that a control over the spin inertia via the nonequilibrium average OAM induced by orbital pumping~\cite{Go2025} offers a promising avenue for confirming OAM as the true origin of spin inertia. As per our analysis, such a change in the average OAM should directly affect the visibility or the dynamic susceptibility of the spin nutation mode while leaving its frequency largely unchanged. Alternately, measuring the effective g-factor of the spin nutation mode by recording its frequency vs.~applied field should offer a strong indication of OAM being at its origin.}

If validated by experiments, our proposed model will provide a microscopic understanding of spin inertia and nutation thereby enabling prediction for a wide range of materials. To this end, however, a determination of Russel-Saunders spin-orbit coupling for the materials will be needed. This could be accomplished either via first-principles calculations or by other experimental measurements.

Our analysis here establishes a direct link between the spin inertia parameter and the net OAM content in the magnet, which is negligibly small in nearly all materials. However, the recently emerged field of orbitronics suggests that this OAM content can potentially become large in driven system. Thus, if OAM is indeed confirmed to be the origin of spin inertia, we envisage an effective control of the spin inertia and nutation by altering the OAM content in the magnetic material via recently discovered orbitronic techniques.

\section*{Acknowledgment}

We thank Philipp Pirro and Ulrich Nowak for valuable discussions. Financial support by the DFG (German Research Foundation) via Spin+X TRR 173-268565370 (project A13) is gratefully acknowledged.


\onecolumngrid

\appendix


\section{Dynamical magnetic susceptibility}
\label{susceptibility}
    \begin{figure*}[tbh!]
	\centering
		\subfloat[][]{\includegraphics[width=.45\textwidth]{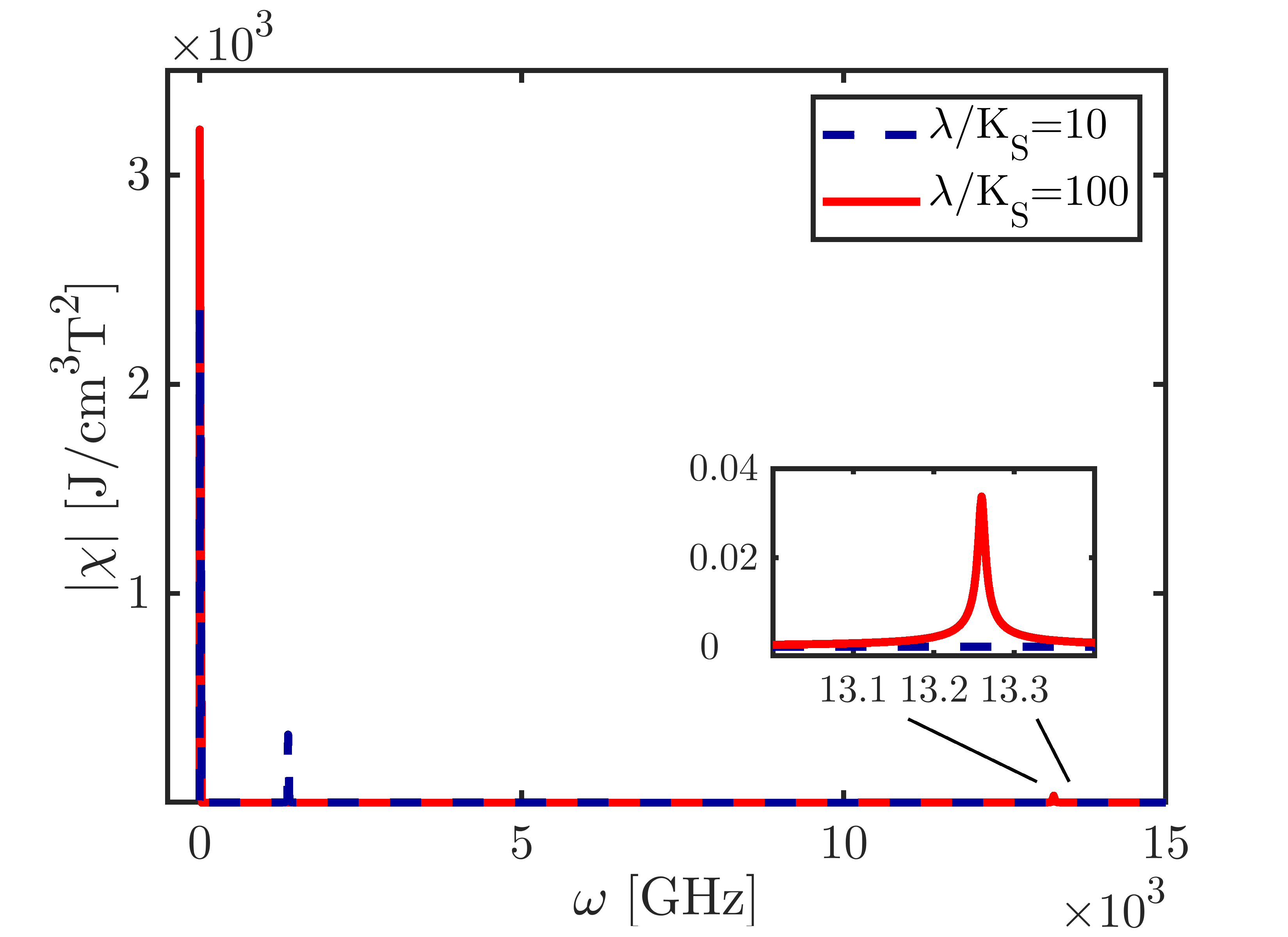}\label{susceptible a}}
		\subfloat[][]{\includegraphics[width=.45\textwidth]{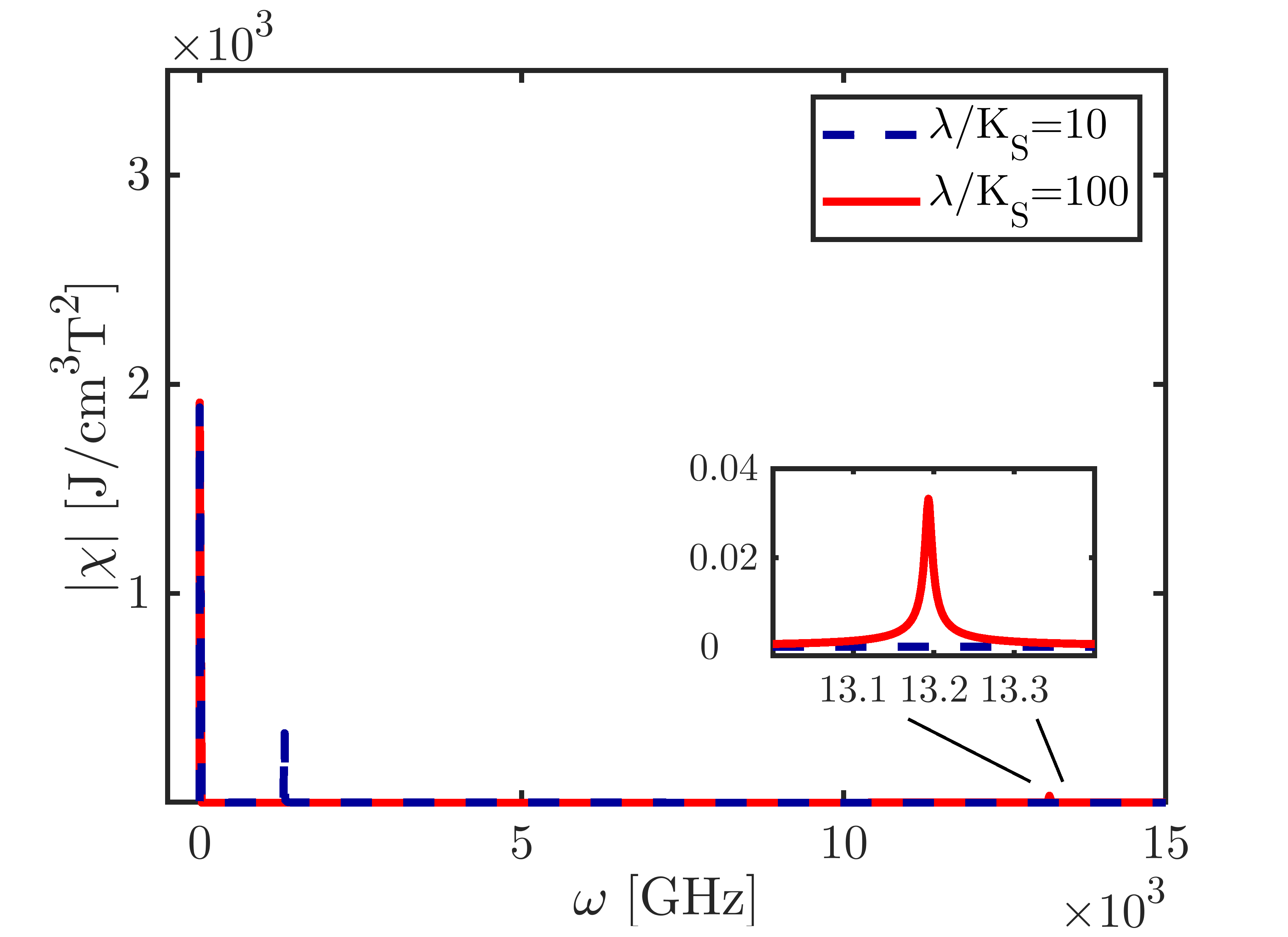}\label{susceptible b}}
		\caption{\label{susceptible} Dynamic susceptibility $\abs{\chi_{c 1}+\chi_{c 2}}$ for (a) ferromagnetic and (b) antiferromagnetic RS coupling for $\lambda/K_S=10$ and $\lambda/K_S=100$.
		The other parameters are $K_S=K_L=2.5\times10^{-2}\ \text{cm$^3\cdot$T$^2\cdot$J}^{-1}$, $M_{S0}=1.5\ \text{J$\cdot$T$^{-1}\cdot$cm$^{-3}$}$, $M_{L0}/M_{S0}=0.1$, $\mu_0H_0/K_SM_{S0}=0.2$, $\abs{\gamma_L}=1.76\times10^{11}\ \text{s$^{-1}\cdot$T}^{-1}$, $\abs{\gamma_S}/\abs{\gamma_L}$=2, $\alpha_{SS}/\abs{\gamma_S}M_{S0}K_S=\alpha_{LL}/\abs{\gamma_L}M_{L0}K_S=0.004$ and $\alpha_{SL}/\abs{\gamma_S}M_{L0}K_S=\alpha_{LS}/\abs{\gamma_L}M_{S0}K_S=0.002$.}
	\end{figure*}
	One reason why spin inertia is a relatively new discovery is that the nutation resonances could not be seen in FMR experiments~\cite{Kittel1948, Farle1998, Polder1949}. With our model this absence of nutation mode in experiments can also be justified.
    
    To calculate our expected result for a FMR experiment we choose the external magnetic field in Eq.~(\ref{freeenergy}) as:
	\begin{equation}
		\va{H}_0=H_0\vu{e}_z+h_0\cos(\omega t)\vu{e}_x,
	\end{equation}
	and calculate the dynamic magnetic susceptibilities ($\chi_{l n}'$) of the eigenmodes. These are the eigenvalues of the susceptibility matrix ${X}_l'$ from the relation
	\begin{equation}
		\mqty(\mathcal{M}_{Sl}\\ \mathcal{M}_{Ll})=X_l'\mu_0\mqty(h_0\\h_0).
	\end{equation}
	For ferromagnetic and antiferromagnetic coupling these eigenvalues are calculated as:
	\begin{align}
		\chi_{F n}'=2\Biggl[&\frac{\Omega_S-\mathfrak{s}_c\omega}{\abs{\gamma_S}M_{S0}}+i\omega\frac{\alpha_{SS}}{\abs{\gamma_S}M_{S0}}+\frac{\Omega_L-\mathfrak{s}_c\omega}{\abs{\gamma_L}M_{L0}}+i\omega\frac{\alpha_{LL}}{\abs{\gamma_L}M_{L0}}\nonumber\\
			&-\mathfrak{s}_a\sqrt{\qty(\frac{\Omega_S-\mathfrak{s}_c\omega}{\abs{\gamma_S}M_{S0}}+i\omega\frac{\alpha_{SS}}{\abs{\gamma_S}M_{S0}}-\frac{\Omega_L-\mathfrak{s}_c\omega}{\abs{\gamma_L}M_{L0}}-i\omega\frac{\alpha_{LL}}{\abs{\gamma_L}M_{L0}})^2+4\qty(\lambda-i\omega\frac{\alpha_{SL}}{\abs{\gamma_S}M_{L0}})^2}\Biggr]^{-1},\\
		\chi_{A n}'=2\Biggl[&\frac{\Omega_S-\mathfrak{s}_c\omega}{\abs{\gamma_S}M_{S0}}+i\omega\frac{\alpha_{SS}}{\abs{\gamma_S}M_{S0}}+\frac{\Omega_L+\mathfrak{s}_c\omega}{\abs{\gamma_L}M_{L0}}+i\omega\frac{\alpha_{LL}}{\abs{\gamma_L}M_{L0}}\nonumber\\
			&+\mathfrak{s}_a\sqrt{\qty(\frac{\Omega_S-\mathfrak{s}_c\omega}{\abs{\gamma_S}M_{S0}}+i\omega\frac{\alpha_{SS}}{\abs{\gamma_S}M_{S0}}-\frac{\Omega_L+\mathfrak{s}_c\omega}{\abs{\gamma_L}M_{L0}}-i\omega\frac{\alpha_{LL}}{\abs{\gamma_L}M_{L0}})^2+4\qty(\lambda+i\omega\frac{\alpha_{SL}}{\abs{\gamma_S}M_{L0}})^2}\Biggr]^{-1}.
	\end{align}

	As magnetization measurements are not usually able to differentiate between the magnetization of the different sublattices we also calculate the eigenvalues of the total susceptibility matrix $X_l$ in the following relation:
	\begin{equation}
		\va{M}=X_l \cos(\omega t)\mu_0h_0\vu{e}_x,\label{eqSus}
	\end{equation}	 
	where $\va{M}=\va{M}_S+\va{M}_L$. We have to find how the sublattices combine in the eigenmodes of $X_l$. To do so, we linearize $\chi'_{\mathcal{L}n}$ in the limit ${M_{S0}}/{M_{L0}}\gg 1$ and therefore 
    \begin{equation}
    \frac{\Omega_L-\mathfrak{s}_c\omega}{\abs{\gamma_L}M_{L0}}+i\omega\frac{\alpha_{SS}}{\abs{\gamma_S}M_{S0}}-\mathfrak{s}_o\frac{\Omega_S-\mathfrak{s}_c\omega}{\abs{\gamma_S}M_{S0}}-\mathfrak{s}_oi\omega\frac{\alpha_{LL}}{\abs{\gamma_L}M_{L0}}>\lambda+i\omega\frac{\alpha_{SL}}{\abs{\gamma_S}M_{L0}},
    \end{equation}
    which allows us to calculate the eigenmodes and find the real susceptibilities as:
	\begin{align}
		\chi_{Fn}=\frac{1}{2}\sum\limits_{c=r,l}\frac{\frac{\Omega_L-\mathfrak{s}_c\omega}{\abs{\gamma_L}M_{L0}}+i\omega\frac{\alpha_{LL}}{\abs{\gamma_L}M_{L0}}-\frac{\Omega_S-\mathfrak{s}_c\omega}{\abs{\gamma_S}M_{S0}}-i\omega\frac{\alpha_{SS}}{\abs{\gamma_S}M_{S0}}+\mathfrak{s}_a\qty(\lambda-i\omega\frac{\alpha_{SL}}{\abs{\gamma_S}M_{L0}})}{\sqrt{\qty(\frac{\Omega_L-\mathfrak{s}_c\omega}{\abs{\gamma_L}M_{L0}}-\frac{\Omega_S-\mathfrak{s}_c\omega}{\abs{\gamma_S}M_{S0}})^2+\qty(\omega\frac{\alpha_{LL}}{\abs{\gamma_L}M_{L0}}-\omega\frac{\alpha_{SS}}{\abs{\gamma_S}M_{S0}})^2+\lambda^2+\qty(\omega\frac{\alpha_{SL}}{\abs{\gamma_S}M_{L0}})^2}}\chi_{Fn}',\\
		\chi_{An}=\frac{1}{2}\sum\limits_{c=r,l}\frac{\frac{\Omega_L+\mathfrak{s}_c\omega}{\abs{\gamma_L}M_{L0}}+i\omega\frac{\alpha_{LL}}{\abs{\gamma_L}M_{L0}}+\frac{\Omega_S-\mathfrak{s}_c\omega}{\abs{\gamma_S}M_{S0}}+i\omega\frac{\alpha_{SS}}{\abs{\gamma_S}M_{S0}}+\mathfrak{s}_a\qty(\lambda+i\omega\frac{\alpha_{SL}}{\abs{\gamma_S}M_{L0}})}{\sqrt{\qty(\frac{\Omega_L+\mathfrak{s}_c\omega}{\abs{\gamma_L}M_{L0}}+\frac{\Omega_S-\mathfrak{s}_c\omega}{\abs{\gamma_S}M_{S0}})^2+\qty(\omega\frac{\alpha_{LL}}{\abs{\gamma_L}M_{L0}}+\omega\frac{\alpha_{SS}}{\abs{\gamma_S}M_{S0}})^2+\lambda^2+\qty(\omega\frac{\alpha_{SL}}{\abs{\gamma_S}M_{L0}})^2}}\chi_{An}',
	\end{align}
	where $\chi_{ln}'$ mostly contain the interaction between the resonance frequency and damping. The corrections in $\chi_{l n}$ meanwhile contain the different coupling of the modes to external fields due to their different projections of the external field $h_0\vu{e}_X$ onto the eigenmodes. The total susceptibility $\chi_l$ is then the sum of the susceptibilities of the two modes $\chi_{l1}$ and $\chi_{l2}$. It is a complex value where the phase determines the phase difference between the magnetization and the driving field, and the norm gives the strength of the magnetization . 

    The dynamic susceptibilities for ferromagnetic and antiferromagnetic coupling are plotted in Fig.~\ref{susceptible} for different values of $\lambda$. We observe that at a particular value of $\lambda$ there are two peaks in the system both for ferromagnetic and antiferromagnetic coupling. As $\lambda$ increases, the difference between the peaks also increases. We attribute the peak at higher frequency to the nutation mode and the one at lower frequency to the precessional mode. 
    
    One can also find that for higher value of $\lambda$, as is the case in typical materials, the peak amplitude for the nutation mode is much smaller than that of the precession mode [This is shown in the inset of Fig. \ref{susceptible a} and \ref{susceptible b}]. So, it becomes quite clear why this peak is not captured in FMR experiments. 
    
	Thus we find that the nutation frequency becomes harder to detect, as for higher RS coupling its amplitude is much less than that of precession mode. Therefore, our model agrees with experimental results.

\section{Detailed calculation of resonance frequencies}\label{Afrequencies}

    To derive Eq.~(\ref{matrix}) from the Eqs.~(\ref{dglS},\ref{dglL},\ref{freeenergy}) we start by evaluating the effective fields $\va{H}_S$ and $\va{H}_L$ with Eq.~(\ref{variation}) as
    \begin{align}
        \mu_0\va{H}_S&=\mu_0{H}_0\vu{e}_z+2K_SM_{S0}\vu{e}_z+\mathfrak{s}_o\lambda\va{M}_L,\\
        \mu_0\va{H}_L&=\mu_0{H}_0\vu{e}_z+\mathfrak{s}_o2K_LM_{L0}\vu{e}_z+\mathfrak{s}_o\lambda\va{M}_S,
    \end{align}
    where we use the small excitation approximation for $M_z\approx M_0$. This gives us the differential equations:
    \begin{align}
        \dot{M}_{Sx}=&-\abs{\gamma_S}M_{Sy}\qty(\mu_0H_0+2K_SM_{S0}+\lambda M_{L0})+\mathfrak{s}_o\lambda \abs{\gamma_S}M_{S0}M_{Ly}-\alpha_{SS}M_{S0}\dot{M}_{Sy}-\alpha_{SL}M_{S0}\dot{M}_{Ly},\\
        \dot{M}_{Sy}=&+\abs{\gamma_S}M_{Sx}\qty(\mu_0H_0+2K_SM_{S0}+\lambda M_{L0})-\mathfrak{s}_o\lambda\abs{\gamma_S} M_{S0}M_{Lx}+\alpha_{SS}M_{S0}\dot{M}_{Sx}+\alpha_{SL}M_{S0}\dot{M}_{Lx},\\
        \dot{M}_{Lx}=&-\abs{\gamma_L}M_{Ly}\qty(\mu_0H_0+2K_LM_{L0}+\mathfrak{s}_o\lambda M_{S0})+\lambda \abs{\gamma_L}M_{L0}M_{Sy}-\mathfrak{s}_o\alpha_{LL}M_{L0}\dot{M}_{Ly}-\mathfrak{s}_o\alpha_{LS}M_{L0}\dot{M}_{Sy},\\
        \dot{M}_{Ly}=&+\abs{\gamma_L}M_{Lx}\qty(\mu_0H_0+2K_LM_{L0}+\mathfrak{s}_o\lambda M_{S0})-\lambda \abs{\gamma_L}M_{L0}M_{Sx}+\mathfrak{s}_o\alpha_{LL}M_{L0}\dot{M}_{Lx}+\mathfrak{s}_o\alpha_{LS}M_{L0}\dot{M}_{Sx}.
    \end{align}
    Here we can define the circular modes 
    \begin{equation}
        M_{\mathcal{L}c}=M_{\mathcal{L}x}+\mathfrak{s}_ciM_{\mathcal{L}y},
    \end{equation}
    where the index $c$ distinguishes between the left rotating mode with $c=l$ and $\mathfrak{s}_l=+1$ and the right rotating mode with $c=r$ and $\mathfrak{s}_r=-1$.
    We can then use the definitions in Eq.~(\ref{BOmega}) and obtain two sets of two coupled equations:
    \begin{align}
        \dot{M}_{Sc}&=\mathfrak{s}_ci\abs{\gamma_S}M_{Sc}\Omega_S-\mathfrak{s}_c\mathfrak{s}_oi\lambda \abs{\gamma_S} M_{S0}M_{Lc}+\mathfrak{s}_ci\alpha_{SS}M_{S0}\dot{M}_{Sc}+\mathfrak{s}_ci\alpha_{SL}M_{S0}\dot{M}_{Lc},\\
        \dot{M}_{Lc}&=\mathfrak{s}_c\mathfrak{s}_oi\abs{\gamma_L}M_{Lc}\Omega_L-\mathfrak{s}_ci\lambda \abs{\gamma_L}M_{L0}M_{Sc}+\mathfrak{s}_c\mathfrak{s}_oi\alpha_{LL}M_{L0}\dot{M}_{Lc}+\mathfrak{s}_c\mathfrak{s}_oi\alpha_{LS}M_{L0}\dot{M}_{Sc}.
    \end{align}
    We can then make the ansatz $M_{\mathcal{L}c}=\mathcal{M}_{\mathcal{L}c}e^{i\nu t}$ with the mode amplitude $\mathcal{M}_{\mathcal{L}c}$ and frequency $\nu$ to write these equation as:
    \begin{align}
       \mathfrak{s}_c\nu\mathcal{M}_{Sc}&=\abs{\gamma_S}\mathcal{M}_{Sc}\Omega_S-\mathfrak{s}_o\lambda \abs{\gamma_S}M_{S0}\mathcal{M}_{Lc}+i\nu\alpha_{SS}{M}_{S0}\mathcal{M}_{Sc}+i\nu\alpha_{SL}M_{S0}\mathcal{M}_{Lc},\\
        \mathfrak{s}_c\nu\mathcal{M}_{Lc}&=\mathfrak{s}_o\abs{\gamma_L}\mathcal{M}_{Lc}\Omega_L-\lambda \abs{\gamma_L}M_{L0}\mathcal{M}_{Sc}+\mathfrak{s}_oi\nu\alpha_{LL}{M}_{L0}\mathcal{M}_{Lc}+\mathfrak{s}_oi\nu\alpha_{LS}M_{L0}\mathcal{M}_{Sc},
    \end{align}
    which can then be written as the matrix equation Eq.~(\ref{matrix})

\section{Deriving sense of rotation from frequencies}\label{rotationsense}
	If we know the eigenfrequency $\omega$ of the circular modes $M_{c}$ where $M_l=\mathcal{M}_le^{i\omega t}$ and $M_r=\mathcal{M}_re^{-i\omega t}$ we can recover the $x$ and $y$ component as:
	\begin{equation}
		M_{x}=\frac{1}{2}\qty(M_{l}+M_{r})\ ,\ M_{y}=\frac{1}{2i}\qty(M_{l}-M_{r}).
	\end{equation}
    In principle the amplitudes $\mathcal{M}_l$ and $\mathcal{M}_l$ are complex quantities but their phase just reflects starting conditions. We will therefore choose them as real.
    
	If we separate these expressions into their real and imaginary parts:
	\begin{align}
		\begin{split}
		M_{x}=&\frac{1}{2}\qty[\mathcal{M}_l\cos(\omega t)+\mathcal{M}_r\cos(\omega t)]
			+\frac{i}{2}\qty[{\mathcal{M}_l}\sin(\omega t)-{\mathcal{M}_r}\sin(\omega t)],
		\end{split}\\
		\begin{split}
		M_{y}=&\frac{1}{2}\qty[{\mathcal{M}_l}\sin(\omega t)+{\mathcal{M}_r}\sin(\omega t)]
		-\frac{i}{2}\qty[{\mathcal{M}_l}\cos(\omega t)-{\mathcal{M}_r}\cos(\omega t)],	
		\end{split}
	\end{align}
	we can use $M_l=M_r^*$ from the definition of the circular basis (where $^*$ denotes the complex conjugate) and our real choice of $\mathcal{M}_l$ and $\mathcal{M}_l$ to find $\mathcal{M}_l=\mathcal{M}_r$, which results in real values for $M_x$ and $M_y$. We can then write the solutions for positive $\omega$ as:
	\begin{equation}
		\va{M}=\mqty(\mathcal{M}_l\cos(\abs{\omega}t)\\\mathcal{M}_l\sin(\abs{\omega}t)\\M_z),
	\end{equation}
	which rotates anticlockwise and for negative $\omega$ as 
	\begin{equation}
		\va{M}=\mqty(\mathcal{M}_l\cos(\abs{\omega}t)\\-\mathcal{M}_l\sin(\abs{\omega}t)\\M_z),
	\end{equation}
	which rotates clockwise.
    
    Note that this result requires the frequency of $M_r$ to be the negative of $M_l$.
 	

\section{Detailed derivation of the single sublattice equation}\label{fullcalc}
	
	From equations \eqref{sdot} and \eqref{ldot}, $\va{M}_S\cp\ddot{\va{M}}_S$ is calculated as:
	\begin{align}
		\va{M}_S\cp\ddot{\va{M}}_S=&-\abs{\gamma_S}\va{M}_S\cp\qty(\dot{\va{M}}_S\cp\mu_0\va{H})\nonumber\\
		&-2\abs{\gamma_S}K_S\va{M}_S\cp\qty(\dot{\va{M}}_S\cp{M}_{Sz}\vu{e}_z+\va{M}_S\cp\dot{M}_{Sz}\vu{e}_z)\nonumber\\
		&-\mathfrak{s}_o\abs{\gamma_S}\lambda\va{M}_S\cp\qty(\dot{\va{M}}_S\cp\va{M}_L+\va{M}_S\cp\dot{\va{M}}_L)
		\\
		=&-\abs{\gamma_S}\mu_0\qty(\va{M}_S\vdot\va{H}_0)\dot{\va{M}}_S\nonumber\\
		&-2\abs{\gamma_S}K_S\qty(\qty({M}_{Sz}\vu{e}_z)^2\dot{\va{M}}_{S}+\va{M}_S\cp\qty(\va{M}_S\cp\dot{M}_{Sz}\vu{e}_z))\nonumber\\
		&-\mathfrak{s}_o\abs{\gamma_S}\lambda\qty(\qty(\va{M}_S\vdot\va{M}_L)\dot{\va{M}}_S+\va{M}_S\cp\qty(\va{M}_S\cp\dot{\va{M}}_L))\label{ddotcross},
	\end{align}
	where we used $\va{M}_S\vdot\dot{\va{M}}_S=0$. Next we can insert Eq.~(\ref{ldot}) into Eq.~(\ref{ddotcross}) to find:
	\begin{align}
		\va{M}_S\cp\ddot{\va{M}}_S=&-\abs{\gamma_S}\mu_0\qty(\va{M}_S\vdot\va{H}_0)\dot{\va{M}}_S\nonumber\\
			&-\abs{\gamma_S}2K_S\qty(\qty({M}_{Sz}\vu{e}_z)^2\dot{\va{M}}_{S}+\va{M}_S\cp\qty(\va{M}_S\cp\dot{M}_{Sz}\vu{e}_z))\nonumber\\
			&-\mathfrak{s}_o\abs{\gamma_S}\lambda\qty(\qty(\va{M}_S\vdot\va{M}_L)\dot{\va{M}}_S)\nonumber\\
			&+\mathfrak{s}_o\abs{\gamma_S}\abs{\gamma_L}\lambda\qty(\va{M}_S\vdot\qty(\mu_0\va{H}_0+K_L{M}_{Lz}\vu{e}_z+\mathfrak{s}_o\lambda\va{M}_S))\va{M}_S\cp\va{M}_L\nonumber\\
			&-\mathfrak{s}_o\abs{\gamma_S}\abs{\gamma_L}\lambda\qty(\va{M}_S\vdot\va{M}_L)\va{M}_S\cp\qty(\mu_0\va{H}_0+K_L{M}_{Lz}\vu{e}_z),
	\end{align}
	which we can be solve for $\va{M}_S\cp\va{M}_L$ and insert into Eq.~(\ref{sdot}) which we can then solve for $\dot{\va{M}}_S$ to find the exact solution:
	\begin{align}
			\dot{\va{M}}_S=&-\frac{\qty(\abs{\gamma_L}\va{M}_S\vdot\qty(\mathfrak{s}_o\lambda\va{M}_{S}+\mu_0\va{H}_0+2K_L{M}_{Lz}\vu{e}_z)+\mathfrak{s}_o\lambda\abs{\gamma_L}\va{M}_S\vdot\va{M}_L)\abs{\gamma_S}\mu_0\va{M}_S\cp\va{H}_0}{\abs{\gamma_L}\va{M}_S\vdot\qty(\mathfrak{s}_o\lambda\va{M}_{S}+\mu_0\va{H}_0+2K_L{M}_{Lz}\vu{e}_z)+\abs{\gamma_S}\va{M}_S\vdot\qty(\mathfrak{s}_o\lambda\va{M}_L+\mu_0\va{H}_0+2K_S{M}_{Sz}\vu{e}_z)}\nonumber\\
			&-\frac{2K_S\abs{\gamma_S}\abs{\gamma_L}\va{M}_S\vdot\qty(\mathfrak{s}_o\lambda\va{M}_{S}+\mu_0\va{H}_0+2K_L{M}_{Lz}\vu{e}_z)\va{M}_S\cp{M}_{Sz}\vu{e}_z}{\abs{\gamma_L}\va{M}_S\vdot\qty(\mathfrak{s}_o\lambda\va{M}_{S}+\mu_0\va{H}_0+2K_L{M}_{Lz}\vu{e}_z)+\abs{\gamma_S}\va{M}_S\vdot\qty(\mathfrak{s}_o\lambda\va{M}_L+\mu_0\va{H}_0+2K_S{M}_{Sz}\vu{e}_z)}\nonumber\\
			&-\mathfrak{s}_o\frac{2\lambda K_L\abs{\gamma_S}\abs{\gamma_L}\qty(\va{M}_S\vdot\va{M}_L)\va{M}_S\cp{M}_{Lz}\vu{e}_z}{\abs{\gamma_L}\va{M}_S\vdot\qty(\mathfrak{s}_o\lambda\va{M}_{S}+\mu_0\va{H}_0+2K_L{M}_{Lz}\vu{e}_z)+\abs{\gamma_S}\va{M}_S\vdot\qty(\mathfrak{s}_o\lambda\va{M}_L+\mu_0\va{H}_0+2K_S{M}_{Sz}\vu{e}_z)}\nonumber\\
			&-\frac{2K_S\abs{\gamma_S}\va{M}_S\cp\qty(\va{M}_S\cp\dot{M}_{Sz}\vu{e}_z)+\va{M}_S\cp\ddot{\va{M}}_S}{\abs{\gamma_L}\va{M}_S\vdot\qty(\mathfrak{s}_o\lambda\va{M}_{S}+\mu_0\va{H}_0+2K_L{M}_{Lz}\vu{e}_z)+\abs{\gamma_S}\va{M}_S\vdot\qty(\mathfrak{s}_o\lambda\va{M}_L+\mu_0\va{H}_0+2K_S{M}_{Sz}\vu{e}_z)}.
	\end{align}
	This Eq. is simplified and has the form of Eq.~(\ref{exact}).

\section{Exploration of near parallel lock of magnetization under strong RS coupling}
    \label{lock}

    As we assume the strong RS coupling to lock $\va{M}_S$ and $\va{M}_L$ near parallel or antiparallel we must be careful to retain $\va{M}_S\cp\va{M}_L\neq0$ to not decouple our sublattices. Accordingly we can only dismiss terms in products of $\va{M}_L$ and $\va{M}_S$ that are smaller than $|\va{M}_S\cp\va{M}_L|$. To determine which terms those are we find it convenient to define a new set of time dependent coordinates with the first basis vector $\vu{e}_S$ parallel to $\va{M}_S$, the second $\vu{e}_\delta$ in the direction of the deviation $\delta$ of $\va{M}_L$ from $\vu{e}_S$ and the third basis vector $\vu{e}_3=\vu{e}_S\cp\vu{e}_\delta$. We can then write the magnetization using the new coordinates and the deviation as:
    \begin{align}
        \va{M}_S=&M_{S0}\vu{e}_S,\\
        \va{M}_L=&\mathfrak{s}_o\sqrt{M_{L0}^2-\delta^2}\vu{e}_S+\delta\vu{e}_\delta.
    \end{align}
    Using these expressions we can easily calculate their cross and scalar products as:
    \begin{align}
        \va{M}_S\cp\va{M}_L=&\delta M_{S0}\vu{e}_3,\\
        \va{M}_S\vdot\va{M}_L=&\mathfrak{s}_oM_{S0}\sqrt{M_{L0}^2+\delta^2}\nonumber\\
            =&\mathfrak{s}_oM_{S0}M_{L0}-\mathfrak{s}_o\frac{M_{S0}}{M_{L0}}\delta^2+\mathcal{O}[\delta^4],
    \end{align}
	which makes it clear that by neglecting all terms of second or higher order in $\delta$ we can dismiss the dynamical parts of the scalar product while retaining the full cross product. 


\section{Detailed expressions of new parameters $\kappa$, $\abs{\gamma_S'}$ and $K_S'$}
\label{lambda}

In this Appendix we present the exact expressions for $\kappa$, $\abs{\gamma_S'}$ and $K_S'$ as well as their approximations including the two highest orders of $\lambda$
\begin{align}
    \kappa=&-\frac{\mathfrak{s}_oM_{S0}}{\abs{\gamma_L}\va{M}_S\vdot\mu_0\va{H}_L+\abs{\gamma_S}\va{M}_S\vdot\mu_0\va{H}_S}\nonumber\\
        \approx&-\frac{\mathfrak{s}_oM_{S0}}{\lambda M_{S0}\qty(\abs{\gamma_L}M_{S0}+\abs{\gamma_S}M_{L0})}+\mathfrak{s}_oM_{S0}\frac{\va{M}_S\vdot\qty(2K_S\abs{\gamma_S}M_{Sz}\vu{e}_z+\mu_0\qty(\abs{\gamma_S}+\abs{\gamma_L})\va{H}_0)}{\lambda^2M_{S0}^2\qty(\abs{\gamma_L}M_{S0}+\mathfrak{s}_o\abs{\gamma_S}M_{L0})^2}\\
    \abs{\gamma_S'}=&\frac{\abs{\gamma_L}\qty(\va{M}_S\vdot\mu_0\va{H}_L+       \mathfrak{s}_o\lambda\va{M}_S\vdot\va{M}_L)}{\abs{\gamma_L}\va{M}_S\vdot\mu_0\va{H}_L+\abs{\gamma_S}\va{M}_S\vdot\mu_0\va{H}_S}\abs{\gamma_S}\nonumber\\
        \approx&\frac{\abs{\gamma_L}\qty(M_{S0}+\mathfrak{s}_oM_{L0})}{\abs{\gamma_L}M_{S0}+\mathfrak{s}_o\abs{\gamma_S}M_{L0}}\abs{\gamma_S}\nonumber\\
        &-\mathfrak{s}_o\frac{2K_S\abs{\gamma_L}^2\qty(M_{S0}+\mathfrak{s}M_{L0})\va{M}_S\vdot M_{Sz}\vu{e}_z+\abs{\gamma_L}\qty(\abs{\gamma_S}M_{S0}+\abs{\gamma_L})M_{L0}\va M_S\vdot\mu_0\va{H}_0}{\lambda M_{S0}\qty(\abs{\gamma_L}M_{S0}+\mathfrak{s}_oM_{L0})^2}\abs{\gamma_S}\\
    K_S'=&\frac{\va{M}_S\vdot\mu_0\va{H}_L}{\va{M}_S\vdot\mu_0\va{H}_L+\mathfrak{s}_o\lambda\va{M}_S\vdot\va{M}_L}K_S\nonumber\\
        \approx&\frac{M_{S0}}{M_{S0}+\mathfrak{s}_oM_{L0}}K_S
        +\frac{M_{S0}\va{M}_S\vdot\mu_0\va{H}_0}{\lambda M_{s0}\qty(M_{S0}+\mathfrak{s}_oM_{L0})^2}K_S
\end{align}

\twocolumngrid
\bibliography{SpinNutation}

@article{Gepraegs2016,
  title = {Origin of the spin Seebeck effect in compensated ferrimagnets},
  volume = {7},
  ISSN = {2041-1723},
  url = {http://dx.doi.org/10.1038/ncomms10452},
  DOI = {10.1038/ncomms10452},
  number = {1},
  journal = {Nature Communications},
  publisher = {Springer Science and Business Media LLC},
  author = {Gepr\"{a}gs,  Stephan and Kehlberger,  Andreas and Coletta,  Francesco Della and Qiu,  Zhiyong and Guo,  Er-Jia and Schulz,  Tomek and Mix,  Christian and Meyer,  Sibylle and Kamra,  Akashdeep and Althammer,  Matthias and Huebl,  Hans and Jakob,  Gerhard and Ohnuma,  Yuichi and Adachi,  Hiroto and Barker,  Joseph and Maekawa,  Sadamichi and Bauer,  Gerrit E. W. and Saitoh,  Eiji and Gross,  Rudolf and Goennenwein,  Sebastian T. B. and Kl\"{a}ui,  Mathias},
  year = {2016},
  pages = {10452},
  month = Feb 
}

@article{Princep2017,
  title = {The full magnon spectrum of yttrium iron garnet},
  volume = {2},
  ISSN = {2397-4648},
  url = {http://dx.doi.org/10.1038/s41535-017-0067-y},
  DOI = {10.1038/s41535-017-0067-y},
  number = {1},
  journal = {npj Quantum Materials},
  publisher = {Springer Science and Business Media LLC},
  author = {Princep,  Andrew J. and Ewings,  Russell A. and Ward,  Simon and Tóth,  Sandor and Dubs,  Carsten and Prabhakaran,  Dharmalingam and Boothroyd,  Andrew T.},
  year = {2017},
  pages = {63},
  month = Nov 
}

@article{Go2025,
  title = {Orbital pumping by magnetization dynamics in ferromagnets},
  author = {Go, Dongwook and Ando, Kazuya and Pezo, Armando and Bl\"ugel, Stefan and Manchon, Aur\'elien and Mokrousov, Yuriy},
  journal = {Phys. Rev. B},
  volume = {111},
  issue = {14},
  pages = {L140409},
  numpages = {7},
  year = {2025},
  month = {Apr},
  publisher = {American Physical Society},
  doi = {10.1103/PhysRevB.111.L140409},
  url = {https://link.aps.org/doi/10.1103/PhysRevB.111.L140409}
}

@article{De2025,
  title = {Magnetic nutation: Transient separation of magnetization from its angular momentum},
  author = {De, Anulekha and Schlegel, Julius and Lentfert, Akira and Scheuer, Laura and Stadtm\"uller, Benjamin and Pirro, Philipp and von Freymann, Georg and Nowak, Ulrich and Aeschlimann, Martin},
  journal = {Phys. Rev. B},
  volume = {111},
  issue = {1},
  pages = {014432},
  numpages = {7},
  year = {2025},
  month = {Jan},
  publisher = {American Physical Society},
  doi = {10.1103/PhysRevB.111.014432},
  url = {https://link.aps.org/doi/10.1103/PhysRevB.111.014432}
}

@article{Juba2019,
  title = {Spin dynamics of $3d$ and $4d$ impurities embedded in prototypical topological insulators},
  author = {Bouaziz, Juba and Dias, Manuel dos Santos and Guimar\~aes, Filipe Souza Mendes and Lounis, Samir},
  journal = {Phys. Rev. Mater.},
  volume = {3},
  issue = {5},
  pages = {054201},
  numpages = {15},
  year = {2019},
  month = {May},
  publisher = {American Physical Society},
  doi = {10.1103/PhysRevMaterials.3.054201},
  url = {https://link.aps.org/doi/10.1103/PhysRevMaterials.3.054201}
}

@misc{Ghosh2026,
      title={Spin Inertia as a Source of Topological Magnons: Chiral Edge States from Coupled Precession and Nutation}, 
      author={Subhadip Ghosh and Mikhail Cherkasskii and Ritwik Mondal and Alexander Mook and Levente Rózsa},
      year={2026},
      eprint={2603.06219},
      archivePrefix={arXiv},
      primaryClass={cond-mat.mtrl-sci},
      url={https://arxiv.org/abs/2603.06219}, 
}

@article{Cherkasskii2024,
  title = {Inertial spin waves in spin spirals},
  author = {Cherkasskii, Mikhail and Mondal, Ritwik and R\'ozsa, Levente},
  journal = {Phys. Rev. B},
  volume = {109},
  issue = {18},
  pages = {184424},
  numpages = {10},
  year = {2024},
  month = {May},
  publisher = {American Physical Society},
  doi = {10.1103/PhysRevB.109.184424},
  url = {https://link.aps.org/doi/10.1103/PhysRevB.109.184424}
}

@article{Titov2022,
  title = {Nutation spin waves in ferromagnets},
  author = {Titov, Sergei V. and Dowling, William J. and Kalmykov, Yuri P. and Cherkasskii, Mikhail},
  journal = {Phys. Rev. B},
  volume = {105},
  issue = {21},
  pages = {214414},
  numpages = {9},
  year = {2022},
  month = {Jun},
  publisher = {American Physical Society},
  doi = {10.1103/PhysRevB.105.214414},
  url = {https://link.aps.org/doi/10.1103/PhysRevB.105.214414}
}

@article{Cherkasskii2021,
  title = {Dispersion relation of nutation surface spin waves in ferromagnets},
  author = {Cherkasskii, Mikhail and Farle, Michael and Semisalova, Anna},
  journal = {Phys. Rev. B},
  volume = {103},
  issue = {17},
  pages = {174435},
  numpages = {6},
  year = {2021},
  month = {May},
  publisher = {American Physical Society},
  doi = {10.1103/PhysRevB.103.174435},
  url = {https://link.aps.org/doi/10.1103/PhysRevB.103.174435}
}

@article{Rodriguez2024,
  title = {Spin Inertia and Auto-Oscillations in Ferromagnets},
  author = {Rodriguez, Rodolfo and Cherkasskii, Mikhail and Jiang, Rundong and Mondal, Ritwik and Etesamirad, Arezoo and Tossounian, Allison and Ivanov, Boris A. and Barsukov, Igor},
  journal = {Phys. Rev. Lett.},
  volume = {132},
  issue = {24},
  pages = {246701},
  numpages = {6},
  year = {2024},
  month = {Jun},
  publisher = {American Physical Society},
  doi = {10.1103/PhysRevLett.132.246701},
  url = {https://link.aps.org/doi/10.1103/PhysRevLett.132.246701}
}

@article{He2024b,
  title = {Influence of the magnetic inertia on the self-oscillation in spin-orbit torque-driven tripartite antiferromagnets with a ${120}^{\ensuremath{\circ}}$ rotation symmetry},
  author = {He, Peng-Bin},
  journal = {Phys. Rev. B},
  volume = {110},
  issue = {6},
  pages = {064411},
  numpages = {14},
  year = {2024},
  month = {Aug},
  publisher = {American Physical Society},
  doi = {10.1103/PhysRevB.110.064411},
  url = {https://link.aps.org/doi/10.1103/PhysRevB.110.064411}
}

@article{He2024,
  title = {Temporal and spatial attenuation of inertial spin waves driven by spin-transfer torques},
  author = {He, Peng-Bin and Cherkasskii, Mikhail},
  journal = {Phys. Rev. B},
  volume = {110},
  issue = {17},
  pages = {174431},
  numpages = {8},
  year = {2024},
  month = {Nov},
  publisher = {American Physical Society},
  doi = {10.1103/PhysRevB.110.174431},
  url = {https://link.aps.org/doi/10.1103/PhysRevB.110.174431}
}

@article{Titov2021b,
 author = {Titov, S. V. and Coffey, W. T. and Kalmykov, Yu. P. and Zarifakis, M. and Titov, A. S.},
 doi = {10.1103/PhysRevB.103.144433},
 issue = {14},
 journal = {Phys. Rev. B},
 month = {Apr},
 numpages = {7},
 pages = {144433},
 publisher = {American Physical Society},
 title = {Inertial magnetization dynamics of ferromagnetic nanoparticles including thermal agitation},
 url = {https://link.aps.org/doi/10.1103/PhysRevB.103.144433},
 volume = {103},
 year = {2021}
}

@article{Jansen2025,
  title = {Dynamically generated spin interactions and nutational spin inertia in normal metal--ferromagnet heterostructures},
  author = {Johnsen, Christian Svingen and Sudb\o{}, Asle},
  journal = {Phys. Rev. B},
  volume = {111},
  issue = {14},
  pages = {144423},
  numpages = {14},
  year = {2025},
  month = {Apr},
  publisher = {American Physical Society},
  doi = {10.1103/PhysRevB.111.144423},
  url = {https://link.aps.org/doi/10.1103/PhysRevB.111.144423}
}

@article{Kachkachi2025,
  title = {{Magnetization nutation in magnetic semiconductors: Effective spin model with anisotropic RKKY exchange interaction}},
  author = {Kachkachi, H.},
  journal = {Phys. Rev. B},
  volume = {111},
  issue = {1},
  pages = {014410},
  numpages = {17},
  year = {2025},
  month = {Jan},
  publisher = {American Physical Society},
  doi = {10.1103/PhysRevB.111.014410},
  url = {https://link.aps.org/doi/10.1103/PhysRevB.111.014410}
  }

@article{Giordano2020,
 author = {Giordano, Stefano and D\'ejardin, Pierre-Michel},
 doi = {10.1103/PhysRevB.102.214406},
 issue = {21},
 journal = {Phys. Rev. B},
 month = {Dec},
 numpages = {13},
 pages = {214406},
 publisher = {American Physical Society},
 title = {Derivation of magnetic inertial effects from the classical mechanics of a circular current loop},
 url = {https://link.aps.org/doi/10.1103/PhysRevB.102.214406},
 volume = {102},
 year = {2020}
}

@article{Bhattacharjee2012,
 author = {Bhattacharjee, Satadeep and Nordstr\"om, Lars and Fransson, Jonas},
 doi = {10.1103/PhysRevLett.108.057204},
 issue = {5},
 journal = {Phys. Rev. Lett.},
 month = {Jan},
 numpages = {5},
 pages = {057204},
 publisher = {American Physical Society},
 title = {Atomistic Spin Dynamic Method with both Damping and Moment of Inertia Effects Included from First Principles},
 url = {https://link.aps.org/doi/10.1103/PhysRevLett.108.057204},
 volume = {108},
 year = {2012}
}

@article{Osorio2025,
  title = {Optically Induced Magnetic Inertia and Magnons from Non-Markovian Extension of the Landau-Lifshitz-Gilbert Equation},
  author = {Reyes-Osorio, Felipe and Nikoli\ifmmode \acute{c}\else \'{c}\fi{}, Branislav K.},
  journal = {Phys. Rev. Lett.},
  volume = {135},
  issue = {24},
  pages = {246701},
  numpages = {10},
  year = {2025},
  month = {Dec},
  publisher = {American Physical Society},
  doi = {10.1103/sl4k-pcvq},
  url = {https://link.aps.org/doi/10.1103/sl4k-pcvq}
}

@article{Saha2026,
  title = {{Unconventional relaxation dynamics in the chiral magnet ${\mathrm{Co}}_{8}{\mathrm{Zn}}_{7}{\mathrm{Mn}}_{5}$: Evidence of inertial effects}},
  author = {Saha, P. and Singh, M. and Babu, P. D. and Patnaik, S.},
  journal = {Phys. Rev. B},
  volume = {113},
  issue = {10},
  pages = {104427},
  numpages = {13},
  year = {2026},
  month = {Mar},
  publisher = {American Physical Society},
  doi = {10.1103/x2jv-dqhb},
  url = {https://link.aps.org/doi/10.1103/x2jv-dqhb}
}

@article{Unikandanunni2022,
 author = {Unikandanunni, Vivek and Medapalli, Rajasekhar and Asa, Marco and Albisetti, Edoardo and Petti, Daniela and Bertacco, Riccardo and Fullerton, Eric E. and Bonetti, Stefano},
 doi = {10.1103/PhysRevLett.129.237201},
 issue = {23},
 journal = {Phys. Rev. Lett.},
 month = {Nov},
 numpages = {6},
 pages = {237201},
 publisher = {American Physical Society},
 title = {Inertial Spin Dynamics in Epitaxial Cobalt Films},
 url = {https://link.aps.org/doi/10.1103/PhysRevLett.129.237201},
 volume = {129},
 year = {2022}
}

@article{Bajaj2024,
 author = {Bajaj, Robin and Lee, Seung-Cheol and Krishnamurthy, H. R. and Bhattacharjee, Satadeep and Jain, Manish},
 doi = {10.1103/PhysRevB.109.214432},
 issue = {21},
 journal = {Phys. Rev. B},
 month = {Jun},
 numpages = {10},
 pages = {214432},
 publisher = {American Physical Society},
 title = {Calculation of Gilbert damping and magnetic moment of inertia using the torque-torque correlation model within an ab initio Wannier framework},
 url = {https://link.aps.org/doi/10.1103/PhysRevB.109.214432},
 volume = {109},
 year = {2024}
}

@article{Thonig2017,
 author = {Thonig, Danny and Eriksson, Olle and Pereiro, Manuel},
 journal = {Scientific reports},
 number = {1},
 pages = {931},
 publisher = {Nature Publishing Group UK London},
 title = {Magnetic moment of inertia within the torque-torque correlation model},
 url = {https://doi.org/10.1038/s41598-017-01081-z},
 volume = {7},
 year = {2017}
}

@book{Chikazumi2009,
  title={Physics of Ferromagnetism 2e},
  author={Chikazumi, S. and Graham, C.D.},
  isbn={9780199564811},
  lccn={2009497292},
  series={International Series of Monographs on Physics},
  url={https://books.google.de/books?id=f8pdPk3RuowC},
  year={2009},
  publisher={OUP Oxford}
}

@article{Kamra2018,
  title = {Gilbert damping phenomenology for two-sublattice magnets},
  author = {Kamra, Akashdeep and Troncoso, Roberto E. and Belzig, Wolfgang and Brataas, Arne},
  journal = {Phys. Rev. B},
  volume = {98},
  issue = {18},
  pages = {184402},
  numpages = {8},
  year = {2018},
  month = {Nov},
  publisher = {American Physical Society},
  doi = {10.1103/PhysRevB.98.184402},
  url = {https://link.aps.org/doi/10.1103/PhysRevB.98.184402}
}

@article{Goodenough1968,
  title = {Spin-Orbit-Coupling Effects in Transition-Metal Compounds},
  author = {Goodenough, John B.},
  journal = {Phys. Rev.},
  volume = {171},
  issue = {2},
  pages = {466--479},
  numpages = {0},
  year = {1968},
  month = {Jul},
  publisher = {American Physical Society},
  doi = {10.1103/PhysRev.171.466},
  url = {https://link.aps.org/doi/10.1103/PhysRev.171.466}
}

@article{Koseki2019,
author = {Koseki, Shiro and Matsunaga, Nikita and Asada, Toshio and Schmidt, Michael and Gordon, Mark},
year = {2019},
month = {02},
pages = {},
title = {Spin-Orbit Coupling Constants in Atoms and Ions of Transition Elements:Comparison of Effective Core Potentials, Model Core Potentials, and All-Electron Methods},
volume = {123},
journal = {The Journal of Physical Chemistry A},
doi = {10.1021/acs.jpca.8b09218}
}

@article{Vivek2022,
  title = {Inertial Spin Dynamics in Epitaxial Cobalt Films},
  author = {Unikandanunni, Vivek and Medapalli, Rajasekhar and Asa, Marco and Albisetti, Edoardo and Petti, Daniela and Bertacco, Riccardo and Fullerton, Eric E. and Bonetti, Stefano},
  journal = {Phys. Rev. Lett.},
  volume = {129},
  issue = {23},
  pages = {237201},
  numpages = {6},
  year = {2022},
  month = {Nov},
  publisher = {American Physical Society},
  doi = {10.1103/PhysRevLett.129.237201},
  url = {https://link.aps.org/doi/10.1103/PhysRevLett.129.237201}
}

@Article{Neeraj2021,
author={Neeraj, Kumar
and Awari, Nilesh
and Kovalev, Sergey
and Polley, Debanjan
and Zhou Hagstr{\"o}m, Nanna
and Arekapudi, Sri Sai Phani Kanth
and Semisalova, Anna
and Lenz, Kilian
and Green, Bertram
and Deinert, Jan-Christoph
and Ilyakov, Igor
and Chen, Min
and Bawatna, Mohammed
and Scalera, Valentino
and d'Aquino, Massimiliano
and Serpico, Claudio
and Hellwig, Olav
and Wegrowe, Jean-Eric
and Gensch, Michael
and Bonetti, Stefano},
title={Inertial spin dynamics in ferromagnets},
journal={Nature Physics},
year={2021},
month={Feb},
day={01},
volume={17},
number={2},
pages={245-250},
issn={1745-2481},
doi={10.1038/s41567-020-01040-y},
url={https://doi.org/10.1038/s41567-020-01040-y}
}

@ARTICLE{Gilbert2004,

  author={Gilbert, T.L.},

  journal={IEEE Transactions on Magnetics}, 

  title={A phenomenological theory of damping in ferromagnetic materials}, 

  year={2004},

  volume={40},

  number={6},

  pages={3443-3449},

  doi={10.1109/TMAG.2004.836740}}

@article{Ciornei2011,
  title = {Magnetization dynamics in the inertial regime: Nutation predicted at short time scales},
  author = {Ciornei, M.-C. and Rub\'{\i}, J. M. and Wegrowe, J.-E.},
  journal = {Phys. Rev. B},
  volume = {83},
  issue = {2},
  pages = {020410},
  numpages = {4},
  year = {2011},
  month = {Jan},
  publisher = {American Physical Society},
  doi = {10.1103/PhysRevB.83.020410},
  url = {https://link.aps.org/doi/10.1103/PhysRevB.83.020410}
}

@article{Mondal2021,
  title = {Spin pumping at terahertz nutation resonances},
  author = {Mondal, Ritwik and Kamra, Akashdeep},
  journal = {Phys. Rev. B},
  volume = {104},
  issue = {21},
  pages = {214426},
  numpages = {7},
  year = {2021},
  month = {Dec},
  publisher = {American Physical Society},
  doi = {10.1103/PhysRevB.104.214426},
  url = {https://link.aps.org/doi/10.1103/PhysRevB.104.214426}
}

@article{Buenemann2008,
  title = {Spin-Orbit Coupling in Ferromagnetic Nickel},
  author = {B\"unemann, J. and Gebhard, F. and Ohm, T. and Weiser, S. and Weber, W.},
  journal = {Phys. Rev. Lett.},
  volume = {101},
  issue = {23},
  pages = {236404},
  numpages = {4},
  year = {2008},
  month = {Dec},
  publisher = {American Physical Society},
  doi = {10.1103/PhysRevLett.101.236404},
  url = {https://link.aps.org/doi/10.1103/PhysRevLett.101.236404}
}

@book{Coey2010, place={Cambridge}, title={Magnetism and Magnetic Materials}, publisher={Cambridge University Press}, author={Coey, J. M. D.}, year={2010}}

@book{Kittel2005,
  title={Introduction to Solid State Physics},
  author={Kittel, Charles},
  edition={8},
  year={2005},
  publisher={Wiley}
}

@book{BransdenJoachain2003,
  title={Physics of Atoms and Molecules},
  author={Bransden, B. H. and Joachain, C. J.},
  edition={2},
  year={2003},
  publisher={Pearson}
}

@article{Liu2011,
author = {Liu, Ying and Liu, Yue and Liu, Bihui},
title = {A New Method for Obtaining Russell-Saunders Terms},
journal = {Journal of Chemical Education},
volume = {88},
number = {3},
pages = {295-298},
year = {2011},
doi = {10.1021/ed100721q},
URL = {https://doi.org/10.1021/ed100721q},
}

@article{Condon1952,
    author = {Condon, E. U. and Shortley, G. H. and Ufford, C. W.},
    title = {The Theory of Atomic Spectra},
    journal = {American Journal of Physics},
    volume = {20},
    number = {6},
    pages = {383-383},
    year = {1952},
    month = {09},
    issn = {0002-9505},
    doi = {10.1119/1.1933256},
    url = {https://doi.org/10.1119/1.1933256}
}

@article{Bagus2008,
title = {A new analysis of X-ray adsorption branching ratios: Use of Russell–Saunders coupling},
journal = {Chemical Physics Letters},
volume = {455},
number = {4},
pages = {331-334},
year = {2008},
issn = {0009-2614},
doi = {https://doi.org/10.1016/j.cplett.2008.02.085},
url = {https://www.sciencedirect.com/science/article/pii/S0009261408002893},
author = {Paul S. Bagus and Hajo Freund and Helmut Kuhlenbeck and Eugene S. Ilton}
}

@article{Laxman2011,
  author  = {M. Lakshmanan},
  title   = {The fascinating world of the Landau–Lifshitz–Gilbert equation: an overview},
  journal = {Philosophical Transactions of the Royal Society A},
  volume  = {369},
  number  = {1939},
  pages   = {1280--1300},
  year    = {2011},
  doi     = {10.1098/rsta.2010.0319},
  url     = {http://doi.org/10.1098/rsta.2010.0319}
}

@article{Mondal2023,
title = {Inertial effects in ultrafast spin dynamics},
journal = {Journal of Magnetism and Magnetic Materials},
volume = {579},
pages = {170830},
year = {2023},
issn = {0304-8853},
doi = {https://doi.org/10.1016/j.jmmm.2023.170830},
url = {https://www.sciencedirect.com/science/article/pii/S0304885323004791},
author = {Ritwik Mondal and Levente Rózsa and Michael Farle and Peter M. Oppeneer and Ulrich Nowak and Mikhail Cherkasskii}
}

@article{Ghosh2024,
  title = {Theory of tensorial magnetic inertia in terahertz spin dynamics},
  author = {Ghosh, Subhadip and Cherkasskii, Mikhail and Barsukov, Igor and Mondal, Ritwik},
  journal = {Phys. Rev. B},
  volume = {110},
  issue = {17},
  pages = {174430},
  numpages = {12},
  year = {2024},
  month = {Nov},
  publisher = {American Physical Society},
  doi = {10.1103/PhysRevB.110.174430},
  url = {https://link.aps.org/doi/10.1103/PhysRevB.110.174430}
}

@article{Titov2021,
  title = {Deterministic inertial dynamics of the magnetization of nanoscale ferromagnets},
  author = {Titov, S. V. and Coffey, W. T. and Kalmykov, Yu. P. and Zarifakis, M.},
  journal = {Phys. Rev. B},
  volume = {103},
  issue = {21},
  pages = {214444},
  numpages = {9},
  year = {2021},
  month = {Jun},
  publisher = {American Physical Society},
  doi = {10.1103/PhysRevB.103.214444},
  url = {https://link.aps.org/doi/10.1103/PhysRevB.103.214444}
}

@article{Mondal2017,
  title = {Relativistic theory of magnetic inertia in ultrafast spin dynamics},
  author = {Mondal, Ritwik and Berritta, Marco and Nandy, Ashis K. and Oppeneer, Peter M.},
  journal = {Phys. Rev. B},
  volume = {96},
  issue = {2},
  pages = {024425},
  numpages = {9},
  year = {2017},
  month = {Jul},
  publisher = {American Physical Society},
  doi = {10.1103/PhysRevB.96.024425},
  url = {https://link.aps.org/doi/10.1103/PhysRevB.96.024425}
}

@article{Cherkasskii2022,
  title = {Theory of inertial spin dynamics in anisotropic ferromagnets},
  author = {Cherkasskii, Mikhail and Barsukov, Igor and Mondal, Ritwik and Farle, Michael and Semisalova, Anna},
  journal = {Phys. Rev. B},
  volume = {106},
  issue = {5},
  pages = {054428},
  numpages = {10},
  year = {2022},
  month = {Aug},
  publisher = {American Physical Society},
  doi = {10.1103/PhysRevB.106.054428},
  url = {https://link.aps.org/doi/10.1103/PhysRevB.106.054428}
}

@article{Dhali2024,
doi = {10.1088/1361-648X/ad353a},
url = {https://doi.org/10.1088/1361-648X/ad353a},
year = {2024},
month = {mar},
publisher = {IOP Publishing},
volume = {36},
number = {25},
pages = {255804},
author = {Dhali, Prasad and Mondal, Ritwik},
title = {Theory of tensorial Gilbert damping in antiferromagnets},
journal = {Journal of Physics: Condensed Matter}
}

@article{Evans2014,
doi = {10.1088/0953-8984/26/10/103202},
url = {https://doi.org/10.1088/0953-8984/26/10/103202},
year = {2014},
month = {feb},
publisher = {IOP Publishing},
volume = {26},
number = {10},
pages = {103202},
author = {Evans, R F L and Fan, W J and Chureemart, P and Ostler, T A and Ellis, M O A and Chantrell, R W},
title = {Atomistic spin model simulations of magnetic nanomaterials},
journal = {Journal of Physics: Condensed Matter}
}

@article{Skubic2008,
doi = {10.1088/0953-8984/20/31/315203},
url = {https://doi.org/10.1088/0953-8984/20/31/315203},
year = {2008},
month = {jul},
publisher = {},
volume = {20},
number = {31},
pages = {315203},
author = {Skubic, B and Hellsvik, J and Nordström, L and Eriksson, O},
title = {A method for atomistic spin dynamics simulations: implementation and examples},
journal = {Journal of Physics: Condensed Matter}
}

@article{Shovon2022,
    author = {Li, Jingwen and Yang, Chia-Jung and Mondal, Ritwik and Tzschaschel, Christian and Pal, Shovon},
    title = {A perspective on nonlinearities in coherent magnetization dynamics},
    journal = {Applied Physics Letters},
    volume = {120},
    number = {5},
    pages = {050501},
    year = {2022},
    month = {02},
    issn = {0003-6951},
    doi = {10.1063/5.0075999},
    url = {https://doi.org/10.1063/5.0075999},
}

@article{Arpita2024,
  title = {Role of material-dependent properties in THz field-derivative-torque-induced nonlinear magnetization dynamics},
  author = {Dutta, Arpita and Mukherjee, Pratyay and Sarangi, Swosti P. and Bhattacharjee, Somasree and Pal, Shovon and Mondal, Ritwik},
  journal = {Phys. Rev. Mater.},
  volume = {8},
  issue = {11},
  pages = {114404},
  numpages = {9},
  year = {2024},
  month = {Nov},
  publisher = {American Physical Society},
  doi = {10.1103/PhysRevMaterials.8.114404},
  url = {https://link.aps.org/doi/10.1103/PhysRevMaterials.8.114404}
}

@article{Mondal2022,
  title = {Inertial spin waves in ferromagnets and antiferromagnets},
  author = {Mondal, Ritwik and R\'ozsa, Levente},
  journal = {Phys. Rev. B},
  volume = {106},
  issue = {13},
  pages = {134422},
  numpages = {14},
  year = {2022},
  month = {Oct},
  publisher = {American Physical Society},
  doi = {10.1103/PhysRevB.106.134422},
  url = {https://link.aps.org/doi/10.1103/PhysRevB.106.134422}
}

@article{Rozsa2013,
doi = {10.1088/0953-8984/25/50/506002},
url = {https://doi.org/10.1088/0953-8984/25/50/506002},
year = {2013},
month = {nov},
publisher = {IOP Publishing},
volume = {25},
number = {50},
pages = {506002},
author = {R\'ozsa, L and Udvardi, L and Szunyogh, L},
title = {Relativistic and thermal effects on the magnon spectrum of a ferromagnetic monolayer},
journal = {Journal of Physics: Condensed Matter}
}

@article{Ghosh2025,
	author={Ghosh, Subhadip and Basu, Darpa Narayan and Mondal, Ritwik},
	title={Engineering spin-wave spectrum via the magnetization inertia tensor},
	journal={Journal of Physics: Condensed Matter},
	url={http://iopscience.iop.org/article/10.1088/1361-648X/ae2a8e},
	year={2025}
}

@article{Boettcher2012,
  title = {Significance of nutation in magnetization dynamics of nanostructures},
  author = {B\"ottcher, D. and Henk, J.},
  journal = {Phys. Rev. B},
  volume = {86},
  issue = {2},
  pages = {020404},
  numpages = {4},
  year = {2012},
  month = {Jul},
  publisher = {American Physical Society},
  doi = {10.1103/PhysRevB.86.020404},
  url = {https://link.aps.org/doi/10.1103/PhysRevB.86.020404}
}

@article{Olive2012,
    author = {Olive, E. and Lansac, Y. and Wegrowe, J.-E.},
    title = {Beyond ferromagnetic resonance: The inertial regime of the magnetization},
    journal = {Applied Physics Letters},
    volume = {100},
    number = {19},
    pages = {192407},
    year = {2012},
    month = {05},
    issn = {0003-6951},
    doi = {10.1063/1.4712056},
    url = {https://doi.org/10.1063/1.4712056},
}

@article{Niemeryer2012,
  title = {Spin Coupling and Orbital Angular Momentum Quenching in Free Iron Clusters},
  author = {Niemeyer, M. and Hirsch, K. and Zamudio-Bayer, V. and Langenberg, A. and Vogel, M. and Kossick, M. and Ebrecht, C. and Egashira, K. and Terasaki, A. and M\"oller, T. and v. Issendorff, B. and Lau, J. T.},
  journal = {Phys. Rev. Lett.},
  volume = {108},
  issue = {5},
  pages = {057201},
  numpages = {5},
  year = {2012},
  month = {Jan},
  publisher = {American Physical Society},
  doi = {10.1103/PhysRevLett.108.057201},
  url = {https://link.aps.org/doi/10.1103/PhysRevLett.108.057201}
}

@article{Kittel1948,
  title = {On the Theory of Ferromagnetic Resonance Absorption},
  author = {Kittel, Charles},
  journal = {Phys. Rev.},
  volume = {73},
  issue = {2},
  pages = {155--161},
  numpages = {0},
  year = {1948},
  month = {Jan},
  publisher = {American Physical Society},
  doi = {10.1103/PhysRev.73.155},
  url = {https://link.aps.org/doi/10.1103/PhysRev.73.155}
}

@article{Farle1998,
doi = {10.1088/0034-4885/61/7/001},
url = {https://doi.org/10.1088/0034-4885/61/7/001},
year = {1998},
month = {jul},
publisher = {},
volume = {61},
number = {7},
pages = {755},
author = {Michael Farle},
title = {Ferromagnetic resonance of ultrathin metallic layers},
journal = {Reports on Progress in Physics},
}

@article{Polder1949,
author = {D. Polder},
title = {VIII. On the theory of ferromagnetic resonance},
journal = {The London, Edinburgh, and Dublin Philosophical Magazine and Journal of Science},
volume = {40},
number = {300},
pages = {99--115},
year = {1949},
publisher = {Taylor \& Francis},
doi = {10.1080/14786444908561215},


URL = { 
    
        https://doi.org/10.1080/14786444908561215
    
    

}
}

@article{Frait1971,
	author = {{Frait, Z.} and {Gemperle, R.}},
	title = {THE g-FACTOR AND SURFACE MAGNETIZATION OF PURE IRON ALONG [100] AND [111] DIRECTIONS},
	DOI= "10.1051/jphyscol:19711182",
	url= "https://doi.org/10.1051/jphyscol:19711182",
	journal = {J. Phys. Colloques},
	year = 1971,
	volume = 32,
	number = C1,
	pages = "C1-541-C1-542",
	month = "",
}

@article{Frait1965,
  title = {Ferromagnetic Resonance in Metals. Frequency Dependence},
  author = {Frait, Zdenek and MacFaden, Harold},
  journal = {Phys. Rev.},
  volume = {139},
  issue = {4A},
  pages = {A1173--A1181},
  numpages = {0},
  year = {1965},
  month = {Aug},
  publisher = {American Physical Society},
  doi = {10.1103/PhysRev.139.A1173},
  url = {https://link.aps.org/doi/10.1103/PhysRev.139.A1173}
}

@book{Rezende2020,
  title={Fundamentals of Magnonics},
  author={Rezende, S.M.},
  isbn={9783030413170},
  series={Lecture Notes in Physics},
  url={https://books.google.de/books?id=zaH0DwAAQBAJ},
  year={2020},
  publisher={Springer International Publishing}
}

@book{Arblaster2018,
  title={Selected Values of the Crystallographic Properties of the Elements},
  author={Arblaster, J.W.},
  isbn={9781627081542},
  lccn={2018933957},
  url={https://books.google.de/books?id=lRBjtgEACAAJ},
  year={2018},
  publisher={ASM International}
}

@article{Prohaska2022,
url = {https://doi.org/10.1515/pac-2019-0603},
title = {Standard atomic weights of the elements 2021 (IUPAC Technical Report)},
title = {},
author = {Thomas Prohaska and Johanna Irrgeher and Jacqueline Benefield and John K. Böhlke and Lesley A. Chesson and Tyler B. Coplen and Tiping Ding and Philip J. H. Dunn and Manfred Gröning and Norman E. Holden and Harro A. J. Meijer and Heiko Moossen and Antonio Possolo and Yoshio Takahashi and Jochen Vogl and Thomas Walczyk and Jun Wang and Michael E. Wieser and Shigekazu Yoneda and Xiang-Kun Zhu and Juris Meija},
pages = {573--600},
volume = {94},
number = {5},
journal = {Pure and Applied Chemistry},
doi = {doi:10.1515/pac-2019-0603},
year = {2022},
lastchecked = {2025-12-17}
}

@article{Scott1962,
  title = {Review of Gyromagnetic Ratio Experiments},
  author = {Scott, G. G.},
  journal = {Rev. Mod. Phys.},
  volume = {34},
  issue = {1},
  pages = {102--109},
  numpages = {0},
  year = {1962},
  month = {Jan},
  publisher = {American Physical Society},
  doi = {10.1103/RevModPhys.34.102},
  url = {https://link.aps.org/doi/10.1103/RevModPhys.34.102}
}

@article{Wallis2006,
    author = {Wallis, T. M. and Moreland, J. and Kabos, P.},
    title = {Einstein–de Haas effect in a NiFe film deposited on a microcantilever},
    journal = {Applied Physics Letters},
    volume = {89},
    number = {12},
    pages = {122502},
    year = {2006},
    month = {09},
    issn = {0003-6951},
    doi = {10.1063/1.2355445},
    url = {https://doi.org/10.1063/1.2355445},
}

@Article{Satoh2017,
author={Satoh, Takuya
and Iida, Ryugo
and Higuchi, Takuya
and Fujii, Yasuhiro
and Koreeda, Akitoshi
and Ueda, Hiroaki
and Shimura, Tsutomu
and Kuroda, Kazuo
and Butrim, V. I.
and Ivanov, B. A.},
title={Excitation of coupled spin--orbit dynamics in cobalt oxide by femtosecond laser pulses},
journal={Nature Communications},
year={2017},
month={Sep},
day={21},
volume={8},
number={1},
pages={638},
issn={2041-1723},
doi={10.1038/s41467-017-00616-2},
url={https://doi.org/10.1038/s41467-017-00616-2}
}

@book{Condon1935,
  title={The Theory of Atomic Spectra},
  author={Condon, E.U. and Shortley, G.H.},
  isbn={9780521047135},
  lccn={35022624},
  url={https://books.google.de/books?id=34o-HU859gIC},
  year={1935},
  publisher={Cambridge University Press}
}

@article{Gomonay2010,
  title = {Spin transfer and current-induced switching in antiferromagnets},
  author = {Gomonay, Helen V. and Loktev, Vadim M.},
  journal = {Phys. Rev. B},
  volume = {81},
  issue = {14},
  pages = {144427},
  numpages = {10},
  year = {2010},
  month = {Apr},
  publisher = {American Physical Society},
  doi = {10.1103/PhysRevB.81.144427},
  url = {https://link.aps.org/doi/10.1103/PhysRevB.81.144427}
}

@incollection{Laughin2014,
title = {19 - Magnetic Properties of Metals and Alloys},
editor = {David E. Laughlin and Kazuhiro Hono},
booktitle = {Physical Metallurgy (Fifth Edition)},
publisher = {Elsevier},
edition = {Fifth Edition},
address = {Oxford},
pages = {1881-2008},
year = {2014},
isbn = {978-0-444-53770-6},
doi = {https://doi.org/10.1016/B978-0-444-53770-6.00019-8},
url = {https://www.sciencedirect.com/science/article/pii/B9780444537706000198},
author = {Michael E. McHenry and David E. Laughlin}
}

@article{Kittel1949,
  title = {On the Gyromagnetic Ratio and Spectroscopic Splitting Factor of Ferromagnetic Substances},
  author = {Kittel, Charles},
  journal = {Phys. Rev.},
  volume = {76},
  issue = {6},
  pages = {743--748},
  numpages = {0},
  year = {1949},
  month = {Sep},
  publisher = {American Physical Society},
  doi = {10.1103/PhysRev.76.743},
  url = {https://link.aps.org/doi/10.1103/PhysRev.76.743}
}

@article{Landau1935,
  title={On the theory of the dispersion of magnetic permeability in ferromagnetic bodies},
  author={Landau, LALE and Lifshitz, Evgeny and others},
  journal={Phys. Z. Sowjetunion},
  volume={8},
  number={153},
  pages={101--114},
  year={1935}
}

@article{Barnett1935,
  title = {Gyromagnetic and Electron-Inertia Effects},
  author = {Barnett, S. J.},
  journal = {Rev. Mod. Phys.},
  volume = {7},
  issue = {2},
  pages = {129--166},
  numpages = {0},
  year = {1935},
  month = {Apr},
  publisher = {American Physical Society},
  doi = {10.1103/RevModPhys.7.129},
  url = {https://link.aps.org/doi/10.1103/RevModPhys.7.129}
}

@article{Edwards2009,
doi = {10.1088/0953-8984/21/14/146002},
url = {https://doi.org/10.1088/0953-8984/21/14/146002},
year = {2009},
month = {mar},
publisher = {},
volume = {21},
number = {14},
pages = {146002},
author = {Edwards, D M and Wessely, O},
title = {The quantum-mechanical basis of an extended Landau–Lifshitz–Gilbert equation for a
current-carrying ferromagnetic wire},
journal = {Journal of Physics: Condensed Matter}
}

@Article{Wieser2015,
author={Wieser, Robert},
title={Description of a dissipative quantum spin dynamics with a Landau-Lifshitz/Gilbert like damping and complete derivation of the classical Landau-Lifshitz equation},
journal={The European Physical Journal B},
year={2015},
month={Mar},
day={25},
volume={88},
number={3},
pages={77},
issn={1434-6036},
doi={10.1140/epjb/e2015-50832-0},
url={https://doi.org/10.1140/epjb/e2015-50832-0}
}

@article{Norambuena2020,
doi = {10.1088/1367-2630/abbbd3},
url = {https://doi.org/10.1088/1367-2630/abbbd3},
year = {2020},
month = {oct},
publisher = {IOP Publishing},
volume = {22},
number = {10},
pages = {103029},
author = {Norambuena, Ariel and Franco, Andrés and Coto, Raúl},
title = {From the open generalized Heisenberg model to the Landau–Lifshitz equation},
journal = {New Journal of Physics}
}

@article{Bernevig2005,
  title = {Orbitronics: The Intrinsic Orbital Current in $p$-Doped Silicon},
  author = {Bernevig, B. Andrei and Hughes, Taylor L. and Zhang, Shou-Cheng},
  journal = {Phys. Rev. Lett.},
  volume = {95},
  issue = {6},
  pages = {066601},
  numpages = {4},
  year = {2005},
  month = {Aug},
  publisher = {American Physical Society},
  doi = {10.1103/PhysRevLett.95.066601},
  url = {https://link.aps.org/doi/10.1103/PhysRevLett.95.066601}
}

@article{Meyer1961,
  title={Experimental g$'$ and g values of Fe, Co, Ni, and their alloys},
  author={Meyer, AJP and Asch, G},
  journal={Journal of Applied Physics},
  volume={32},
  number={3},
  pages={S330--S333},
  year={1961},
  publisher={American Institute of Physics}
}

@article{Wang2025,
author = {Wang, Ping and Chen, Feng and Yang, Yuhe and Hu, Shuai and Li, Yue and Wang, Wenhong and Zhang, Delin and Jiang, Yong},
title = {Orbitronics: Mechanisms, Materials and Devices},
journal = {Advanced Electronic Materials},
volume = {11},
number = {5},
pages = {2400554},
doi = {https://doi.org/10.1002/aelm.202400554},
url = {https://advanced.onlinelibrary.wiley.com/doi/abs/10.1002/aelm.202400554},
year = {2025}
}

@Article{Jo2024,
author={Jo, Daegeun
and Go, Dongwook
and Choi, Gyung-Min
and Lee, Hyun-Woo},
title={Spintronics meets orbitronics: Emergence of orbital angular momentum in solids},
journal={npj Spintronics},
year={2024},
month={Jun},
day={28},
volume={2},
number={1},
pages={19},
issn={2948-2119},
doi={10.1038/s44306-024-00023-6},
url={https://doi.org/10.1038/s44306-024-00023-6}
}

@article{Go2020,
  title = {Orbital torque: Torque generation by orbital current injection},
  author = {Go, Dongwook and Lee, Hyun-Woo},
  journal = {Phys. Rev. Res.},
  volume = {2},
  issue = {1},
  pages = {013177},
  numpages = {12},
  year = {2020},
  month = {Feb},
  publisher = {American Physical Society},
  doi = {10.1103/PhysRevResearch.2.013177},
  url = {https://link.aps.org/doi/10.1103/PhysRevResearch.2.013177}
}

@Article{Lee2021,
author={Lee, Soogil
and Kang, Min-Gu
and Go, Dongwook
and Kim, Dohyoung
and Kang, Jun-Ho
and Lee, Taekhyeon
and Lee, Geun-Hee
and Kang, Jaimin
and Lee, Nyun Jong
and Mokrousov, Yuriy
and Kim, Sanghoon
and Kim, Kab-Jin
and Lee, Kyung-Jin
and Park, Byong-Guk},
title={Efficient conversion of orbital Hall current to spin current for spin-orbit torque switching},
journal={Communications Physics},
year={2021},
month={Nov},
day={01},
volume={4},
number={1},
pages={234},
issn={2399-3650},
doi={10.1038/s42005-021-00737-7},
url={https://doi.org/10.1038/s42005-021-00737-7}
}

@article{Rana2022,
doi = {10.1088/1361-6528/ac2e75},
url = {https://doi.org/10.1088/1361-6528/ac2e75},
year = {2021},
month = {nov},
publisher = {IOP Publishing},
volume = {33},
number = {6},
pages = {062007},
author = {Rana, Bivas and Mondal, Amrit Kumar and Bandyopadhyay, Supriyo and Barman, Anjan},
title = {Applications of nanomagnets as dynamical systems: I},
journal = {Nanotechnology}
}

@article{Barman2020,
    author = {Barman, Anjan and Mondal, Sucheta and Sahoo, Sourav and De, Anulekha},
    title = {Magnetization dynamics of nanoscale magnetic materials: A perspective},
    journal = {Journal of Applied Physics},
    volume = {128},
    number = {17},
    pages = {170901},
    year = {2020},
    month = {11},
    issn = {0021-8979},
    doi = {10.1063/5.0023993},
    url = {https://doi.org/10.1063/5.0023993},
}

@article{Biziere2010,
title = {Switching field modulation of a nanomagnet by resonant microwave spin torque},
journal = {Journal of Magnetism and Magnetic Materials},
volume = {322},
number = {21},
pages = {3320-3323},
year = {2010},
issn = {0304-8853},
doi = {https://doi.org/10.1016/j.jmmm.2010.06.016},
url = {https://www.sciencedirect.com/science/article/pii/S0304885310003975},
author = {N. Biziere and E. Murè and J-Ph. Ansermet},
}

@article{Tay2013,
  title = {Theory of atomistic simulation of spin-transfer torque in nanomagnets},
  author = {Tay, Tiamhock and Sham, L. J.},
  journal = {Phys. Rev. B},
  volume = {87},
  issue = {17},
  pages = {174407},
  numpages = {11},
  year = {2013},
  month = {May},
  publisher = {American Physical Society},
  doi = {10.1103/PhysRevB.87.174407},
  url = {https://link.aps.org/doi/10.1103/PhysRevB.87.174407}
}

@article{Usadel2006,
  title = {Temperature-dependent dynamical behavior of nanoparticles as probed by ferromagnetic resonance using Landau-Lifshitz-Gilbert dynamics in a classical spin model},
  author = {Usadel, K. D.},
  journal = {Phys. Rev. B},
  volume = {73},
  issue = {21},
  pages = {212405},
  numpages = {4},
  year = {2006},
  month = {Jun},
  publisher = {American Physical Society},
  doi = {10.1103/PhysRevB.73.212405},
  url = {https://link.aps.org/doi/10.1103/PhysRevB.73.212405}
}

@article{Wegrowe2012,
    author = {Wegrowe, J.-E. and Ciornei, M.-C.},
    title = {Magnetization dynamics, gyromagnetic relation, and inertial effects},
    journal = {American Journal of Physics},
    volume = {80},
    number = {7},
    pages = {607-611},
    year = {2012},
    month = {07},
    issn = {0002-9505},
    doi = {10.1119/1.4709188},
    url = {https://doi.org/10.1119/1.4709188},
}

@article{Quarenta2024,
  title = {Bath-Induced Spin Inertia},
  author = {Quarenta, Mario Gaspar and Tharmalingam, Mithuss and Ludwig, Tim and Yuan, H. Y. and Karwacki, Lukasz and Verstraten, Robin C. and Duine, Rembert A.},
  journal = {Phys. Rev. Lett.},
  volume = {133},
  issue = {13},
  pages = {136701},
  numpages = {7},
  year = {2024},
  month = {Sep},
  publisher = {American Physical Society},
  doi = {10.1103/PhysRevLett.133.136701},
  url = {https://link.aps.org/doi/10.1103/PhysRevLett.133.136701}
}

@article{Bose2011,
title = {Lagrangian approach and dissipative magnetic systems},
journal = {Physics Letters A},
volume = {375},
number = {24},
pages = {2452-2455},
year = {2011},
issn = {0375-9601},
doi = {https://doi.org/10.1016/j.physleta.2011.05.019},
url = {https://www.sciencedirect.com/science/article/pii/S0375960111005810},
author = {Thomas Bose and Steffen Trimper}
}

@article{Lakshmanan2011,
    author = {Lakshmanan, M.},
    title = {The fascinating world of the Landau–Lifshitz–Gilbert equation: an overview},
    journal = {Philosophical Transactions of the Royal Society A: Mathematical, Physical and Engineering Sciences},
    volume = {369},
    number = {1939},
    pages = {1280-1300},
    year = {2011},
    month = {03},
    issn = {1364-503X},
    doi = {10.1098/rsta.2010.0319},
    url = {https://doi.org/10.1098/rsta.2010.0319},
}

@article{Dutta2025,
author = {Dutta, Arpita and Tzschaschel, Christian and Priyadarshi, Debankit and Mikuni, Kouki and Satoh, Takuya and Mondal, Ritwik and Pal, Shovon},
title = {Evidence of Relativistic Field-Derivative Torque in Nonlinear THz Response of Magnetization Dynamics},
journal = {Advanced Functional Materials},
volume = {35},
number = {7},
pages = {2414582},
doi = {https://doi.org/10.1002/adfm.202414582},
url = {https://advanced.onlinelibrary.wiley.com/doi/abs/10.1002/adfm.202414582},
year = {2025}
}

@article{Bajpai2019,
  title = {Time-retarded damping and magnetic inertia in the Landau-Lifshitz-Gilbert equation self-consistently coupled to electronic time-dependent nonequilibrium Green functions},
  author = {Bajpai, Utkarsh and Nikoli\ifmmode \acute{c}\else \'{c}\fi{}, Branislav K.},
  journal = {Phys. Rev. B},
  volume = {99},
  issue = {13},
  pages = {134409},
  numpages = {11},
  year = {2019},
  month = {Apr},
  publisher = {American Physical Society},
  doi = {10.1103/PhysRevB.99.134409},
  url = {https://link.aps.org/doi/10.1103/PhysRevB.99.134409}
}

@article{Go2021,
doi = {10.1209/0295-5075/ac2653},
url = {https://doi.org/10.1209/0295-5075/ac2653},
year = {2021},
month = {sep},
publisher = {EDP Sciences, IOP Publishing and Società Italiana di Fisica},
volume = {135},
number = {3},
pages = {37001},
author = {Go, Dongwook and Jo, Daegeun and Lee, Hyun-Woo and Kläui, Mathias and Mokrousov, Yuriy},
title = {Orbitronics: Orbital currents in solids},
journal = {Europhysics Letters}
}
\end{document}